\documentclass[12pt,a4paper]{article}
\usepackage{amsmath,amssymb}
\usepackage[dvips]{graphicx}

\begin{document}

\author{R. B. Nevzorov${}^{\dag,\ddag}$, K. A. Ter--Martirosyan${}^{\dag}$, and M. A. Trusov${}^{\dag}$ \\[5mm]
{\itshape ${}^{\dag}$ITEP, Moscow, Russia} \\
{\itshape ${}^{\ddag}$DESY Theory, Hamburg, Germany}}

\title{Renormalization of coupling constants in the minimal SUSY models}

\maketitle

\begin{abstract}
\noindent The considerable part of the parameter space in the MSSM
corresponding to the infrared quasi fixed point scenario is
excluded by LEP\,II bounds on the lightest Higgs boson mass. In
the NMSSM the mass of the lightest Higgs boson reaches its maximum
value in the strong Yukawa coupling limit when Yukawa couplings
are essentially larger than gauge ones at the Grand Unification
scale. In this case the renormalization group flow of Yukawa
couplings and soft SUSY breaking terms is investigated. The
quasi--fixed and invariant lines and surfaces are briefly
discussed. The coordinates of the quasi--fixed points, where all
solutions are concentrated, are given.
\end{abstract}

\newpage

\section{Introduction}

The search for the Higgs boson remains one of the top priorities
for existing accelerators as well as for those still at the design
stage. This is because this boson plays a key role in the Standard
Model which describes all currently available experimental data
with a high degree of accuracy. As a result of the spontaneous
symmetry breaking $SU(2)\otimes U(1)$ the Higgs scalar acquires a
nonzero vacuum expectation value without destroying the Lorentz
invariance, and generates the masses of all fermions and vector
bosons. An analysis of the experimental data using the Standard
Model has shown that there is a $95\%$ probability that its mass
will not exceed $210\text{~GeV}$ \cite{A1}. At the same time,
assuming that there are no new fields and interactions and also no
Landau pole in the solution of the renormalisation group equations
for the self-action constant of Higgs fields up to the scale
$M_\text{Pl}\approx 2.4\cdot 10^{18}\text{~GeV}$, we can show that
$m_h<180\text{~GeV}$ \cite{A2},\cite{A3}. In this case, physical
vacuum is only stable provided that the mass of the Higgs boson is
greater than $135\text{~GeV}$ \cite{A2}-\cite{A6}. However, it
should be noted that this simplified model does not lead to
unification of the gauge constants \cite{A7} and a solution of the
hierarchy problem \cite{A8}. As a result, the construction of a
realistic theory which combines all the fields and interactions is
extremely difficult in this case.

Unification of the gauge constants occurs naturally on the  scale
$M_X\approx 3\cdot 10^{16}\text{~GeV}$ within the supersymmetric
generalisation of the Standard Model, i.e., the Minimal
Supersymmetric Standard Model (MSSM) \cite{A7}. In order that all
the fundamental fermions acquire mass in the MSSM, not one but two
Higgs doublets $H_1$ and $H_2$ must be introduced in the theory,
each acquiring the nonzero vacuum expectation value $v_1$ and
$v_2$ where $v^2=v_1^2+v_2^2=(246\text{~GeV})^2$. The spectrum of
the Higgs sector of the MSSM contains four massive states: two
CP--even, one CP--odd, and one charged. An important
distinguishing feature of the supersymmetric model is the existing
of a light Higgs boson in the CP--even sector. The upper bound on
its mass is determined to a considerable extent by the value
$\tan\beta=v_2/v_1$. In the tree-level approximation the mass of
the lightest Higgs boson in the MSSM does not exceed the mass of
the Z-boson ($M_Z\approx 91.2\text{~GeV}$): $m_h\le M_Z|\cos
2\beta|$ \cite{A9}. Allowance for the contribution of loop
corrections to the effective interaction potential of the Higgs
fields from a $t$--quark and its superpartners significantly
raises the upper bound on its mass:
\begin{equation}
m_h\le\sqrt{M_Z^2\cos^2 2\beta+\Delta}\, . \label{A1}
\end{equation}
Here $\Delta$ are the loop corrections \cite{A10},\cite{A11}. The
values of these corrections are proportional to $m_t^4$, where
$m_t$ is the running mass of $t$--quark which depends
logarithmically on the supersymmetry breaking scale $M_S$ and is
almost independent of the choice of $\tan\beta$. In
\cite{A3},\cite{A5},\cite{A6} bounds on the mass of the Higgs
boson were compared in the Minimal Standard and Supersymmetric
models. The upper bound on the mass of the light CP--even Higgs
boson in the MSSM increases with increasing $\tan\beta$ and for
$\tan\beta\gg 1$ in realistic supersymmetric models with $M_S\le
1000\text{~GeV}$ reaches $125-128\text{~GeV}$.

However, a considerable fraction of the solutions of the system of
MSSM renormalisation group equations is focused near the infrared
quasi-fixed point at $\tan\beta\sim 1$. In the region of parameter
space of interest to us ($\tan\beta\ll 50$) the Yukawa constants
of a $b$--quark ($h_b$) and a $\tau$--lepton ($h_\tau$) are
negligible so that an exact analytic solution can be obtained for
the one--loop renormalisation group equations \cite{A12}. For the
Yukawa constants of a $t$--quark $h_t(t)$ and the gauge constants
$g_i(t)$ its solution has the following form:
\begin{equation}
\begin{gathered}
Y_t(t)=\frac{\dfrac{E(t)}{6F(t)}}{1+\dfrac{1}{6Y_t(0)F(t)}},\quad
\tilde{\alpha}_i(t)=\frac{\tilde{\alpha}_i(0)}{1+b_i\tilde{\alpha}_i(0)t},
\\
E(t)=\left[\frac{\tilde{\alpha}_3(t)}{\tilde{\alpha}_3(0)}\right]^{16/9}
\left[\frac{\tilde{\alpha}_2(t)}{\tilde{\alpha}_2(0)}\right]^{-3}
\left[\frac{\tilde{\alpha}_1(t)}{\tilde{\alpha}_1(0)}\right]^{-13/99}
,\quad F(t)=\int\limits_0^t E(t')dt',
\end{gathered}
\label{A2}
\end{equation}
where the index $i$ has values between 1 and 3,
\begin{gather*}
b_1=33/5,\quad b_2=1,\quad b_3=-3\\
\tilde\alpha_i(t)=\left(\frac{g_i(t)}{4\pi}\right)^2,\quad
Y_i(t)=\left(\frac{h_t(t)}{4\pi}\right)^2.
\end{gather*}
The variable $t$ is determined by a standard method
$t=\ln(M_X^2/q^2)$. The boundary conditions for the
renormalisation group equations are usually set at the grand
unification scale $M_X$ ($t=0$) where the values of all three
Yukawa constants are the same:
$\tilde\alpha_1(0)=\tilde\alpha_2(0)=\tilde\alpha_3(0)=\tilde\alpha_0.$
On the electroweak scale where $h_t^2(0)\gg 1$ the second term in
the denominator of the expression describing the evolution of
$Y_t(t)$ is much smaller than unity and all the solutions are
concentrated in a narrow interval near the quasi-fixed point
$Y_\text{QFP}(t)=E(t)/6F(t)$ \cite{A13}. In other words in the
low-energy range the dependence of $Y_t(t)$ on the initial
conditions on the scale $M_X$ disappears. In addition to the
Yukawa constant of the $t$--quark, the corresponding trilinear
interaction constant of the scalar fields $A_t$ and the
combination of the scalar masses
$\mathfrak{M}^2_t=m_Q^2+m_U^2+m_2^2$ also cease to depend on
$A_t(0)$ and $\mathfrak{M}_t^2(0)$ as $Y_t(0)$ increases. Then on
the electroweak scale near the infrared quasi--fixed point
$A_t(t)$ and $\mathfrak{M}_t^2(t)$ are only expressed in terms of
the gaugino mass on the Grand Unification scale. Formally this
type of solution can be obtained if $Y_t(0)$ is made to go to
infinity. Deviations from this solution are determined by ratio
$1/6F(t)Y_t(0)$ which is of the order of $1/10h_t^2(0)$ on the
electroweak scale.

The properties of the solutions of the system of MSSM
renormalisation group equations and also the particle spectrum
near the infrared quasi-fixed point for $\tan\beta\sim 1$ have
been studied by many authors \cite{A14},\cite{A15}. Recent
investigations \cite{A15}-\cite{A17} have shown that for solutions
$Y_t(t)$ corresponding to the quasi-fixed point regime the value
of $\tan\beta$ is between $1.3$ and $1.8$. These comparatively low
values of $\tan\beta$ yield significantly more stringent bounds on
the mass of the lightest Higgs boson. The weak dependence of the
soft supersymmetry breaking parameters $A_t(t)$ and
$\mathfrak{M}_t^2(t)$ on the boundary conditions near the
quasi-fixed point means that the upper bound on its mass can be
calculated fairly accurately. A theoretical analysis made in
\cite{A15},\cite{A16} showed that $m_h$ does not exceed $94\pm
5\text{~GeV}$. This bound is $25-30\text{~GeV}$ below the absolute
upper bound in the Minimal Supersymmetric Model. Since the lower
bound on the mass of the Higgs boson from LEP\,II data is
$113\text{~GeV}$ \cite{A1}, which for the spectrum of heavy
supersymmetric particles is the same as the corresponding bound on
the mass of the Higgs boson in the Standard Model, a considerable
fraction of the solutions which come out to a quasi--fixed point
in the MSSM, are almost eliminated by existing experimental data.
This provides the stimulus for theoretical analyses of the Higgs
sector in more complex supersymmetric models.

The simplest expansion of the MSSM which can conserve the
unification of the gauge constants and raise the upper bound on
the mass of the lightest Higgs boson is the Next--to--Minimal
Supersymmetric Standard Model (NMSSM) \cite{A18}-\cite{A20}. By
definition the superpotential of the NMSSM is invariant with
respect to the discrete transformations $y'_\alpha=e^{2\pi
i/3}y_\alpha$ of the $Z_3$ group \cite{A19} which means that we
can avoid the problem of the $\mu$-term in supergravity models.
The thing is that the fundamental parameter $\mu$ should be of the
order of $M_\text{Pl}$ since this scale is the only dimensional
parameter characterising the hidden (gravity) sector of the
theory. In this case, however, the Higgs bosons $H_1$ and $H_2$
acquire an enormous mass $m^2_{H_1,H_2}\sim \mu^2\sim
M^2_\text{Pl}$ and no breaking of $SU(2)\otimes U(1)$ symmetry
occurs. In the NMSSM the term $\mu(\hat{H}_1\hat{H}_2)$ in the
superpotential is not invariant with respect to discrete
transformations of the $Z_3$ group and for this reason should be
eliminated from the analysis ($\mu=0$). As a result of the
multiplicative nature of the renormalisation of this parameter,
the term $\mu(q)$ remains zero on any scale $q\le M_X\div
M_\text{Pl}$. However, the absence of mixing of the Higgs doublets
on electroweak scale has the result that $H_1$ acquires no vacuum
expectation value as a result of the spontaneous symmetry breaking
and $d$--type quarks and charged leptons remain massless. In order
to ensure that all quarks and charged leptons acquire nonzero
masses, an additional singlet superfield $\hat{Y}$ with respect to
gauge $SU(2)\otimes U(1)$ transformations is introduced in the
NMSSM. The superpotential of the Higgs sector of the Nonminimal
Supersymmetric Model \cite{A18}-\cite{A20} has the following form:
\begin{equation}
W_h=\lambda
\hat{Y}(\hat{H}_1\hat{H}_2)+\frac{\varkappa}{3}\hat{Y}^3.
\label{A4}
\end{equation}
As a result of the spontaneous breaking of $SU(2)\otimes U(1)$
symmetry, the field $Y$ acquires a vacuum expectation value
($\langle Y\rangle=y/\sqrt{2}$) and the effective $\mu$-term
($\mu=\lambda y/\sqrt{2}$) is generated.

In addition to the Yukawa constants $\lambda$ and $\varkappa$, and
also the Standard Model constants, the Nonminimal Supersymmetric
Model contains a large number of unknown parameters. These are the
so-called soft supersymmetry breaking parameters which are
required to obtain an acceptable spectrum of superpartners of
observable particles form the phenomenological point of view. The
hypothesis on the universal nature of these constants on the Grand
Unification scale allows us to reduce their number in the NMSSM to
three: the mass of all the scalar particles $m_0$, the gaugino
mass $M_{1/2}$, and the trilinear interaction constant of the
scalar fields $A$. In order to avoid strong CP--violation and also
spontaneous breaking of gauge symmetry at high energies
($M_\text{Pl}\gg E\gg m_t$) as a result of which the scalar
superpartners of leptons and quarks would require nonzero vacuum
expectation values, the complex phases of the soft supersymmetry
breaking parameters are assumed to be zero and only positive
values of $m_0^2$ are considered. Naturally universal
supersymmetry breaking parameters appear in the minimal
supergravity model \cite{A31} and also in various string models
\cite{A29},\cite{A32}. In the low-energy region the hypothesis of
universal fundamental parameters allows to avoid the appearance of
neutral currents with flavour changes and can simplify the
analysis of the particle spectrum as far as possible. The
fundamental parameters thus determined on the Grand Unification
scale should be considered as boundary conditions for the system
of renormalisation group equations which describes the evolution
of these constants as far as the electroweak scale or the
supersymmetry breaking scale. The complete system of the
renormalisation group equations of the Nonminimal Supersymmetric
Model can be found in \cite{A33}, \cite{A34}. These experimental
data impose various constraints on the NMSSM parameter space which
were analysed in \cite{A35},\cite{A36}.

The introduction of the neutral field $Y$  in the NMSSM potential
leads to the appearance of a corresponding $F$--term in the
interaction potential of the Higgs fields. As a consequence, the
upper bound on the mass of the lightest Higgs boson is increased:
\begin{equation}
m_h\le\sqrt{\frac{\lambda^2}{2}v^2\sin^2 2\beta+M_Z^2\cos^2
2\beta+\Delta^{(1)}_{11}+\Delta^{(2)}_{11}} . \label{A5}
\end{equation}
The relationship (\ref{A5}) was obtained in the tree-level
approximation ($\Delta_{11}=0$) in \cite{A20}. However, loop
corrections to the effective interaction potential of the Higgs
fields from the $t$--quark and its superpartners play a very
significant role. In terms of absolute value their contribution to
the upper bound on the mass of the Higgs boson remains
approximately the same as in the Minimal Supersymmetric Model.
When calculating the corrections $\Delta_{11}^{(1)}$ and
$\Delta_{11}^{(2)}$ we need to replace the parameter $\mu$ by
$\lambda y/\sqrt{2}$. Studies of the Higgs sector in the
Nonminimal Supersymmetric model and the one--loop corrections to
it were reported in \cite{A24},\cite{A33},\cite{A36}-\cite{A39}.
In \cite{A6} the upper bound on the mass of the lightest Higgs
boson in the NMSSM was compared with the corresponding bounds on
$m_h$ in the Minimal Standard and Supersymmetric Models. The
possibility of a spontaneous CP--violation in the Higgs sector of
the NMSSM was studied in \cite{A39},\cite{A40}.

It follows from condition (\ref{A5}) that the upper bound on $m_h$
increases as $\lambda$ increases. Moreover, it only differs
substantially from the corresponding bound in the MSSM in the
range of small $\tan\beta$. For high values ($\tan\beta\gg 1$) the
value of $\sin 2\beta$ tends to zero and the upper bounds on the
mass of the lightest Higgs boson in the MSSM and NMSSM are almost
the same. The case of small $\tan\beta$ is only achieved for
fairly high values of the Yukawa constant of a $t$--quark $h_t$ on
the electroweak scale ($h_t(t_0)\ge 1$ where
$t_0=\ln(M_X^2/m_t^2)$), and $\tan\beta$ decreases with increasing
$h_t(t_0)$. However, an analysis of the renormalisation group
equations in the NMSSM shows that an increase of the Yukawa
constants on the electroweak scale is accompanied by an increase
of $h_t(0)$ and $\lambda(0)$ on the Grand Unification scale. It
thus becomes obvious that the upper bound on the mass of the
lightest Higgs boson in the Nonminimal Supersymmetric model
reaches its maximum on the strong Yukawa coupling limit, i.e.,
when $h_t(0)\gg g_i(0)$ and $\lambda(0)\gg g_i(0)$.

\section{Renormalization of the Yukawa couplings}

From the point of view of a renormalisation group analysis,
investigation of the NMSSM presents a much more complicated
problem than investigation of the minimal SUSY model. The full set
of renormalization group equations within the NMSSM can be found
in \cite{A33},\cite{A34}. Even in the one--loop approximation,
this set of equations is nonlinear and its analytic solution does
not exist. All equations forming this set can be partitioned into
two groups. The first one contains equations that describe the
evolution of gauge and Yukawa coupling constants, while the second
one includes equations for the parameters of a soft breakdown of
SUSY, which are necessary for obtaining a phenomenologically
acceptable spectrum of superpartners of observable particles.
Since boundary conditions for three Yukawa coupling constants are
unknown, it is very difficult to perform a numerical analysis of
the equations belonging to the first group and of the full set of
the equations given above. In the regime of strong Yukawa
coupling, however, solutions to the renormalisation group
equations are concentrated in a narrow region of the parameter
space near the electroweak scale, and this considerably simplifies
the analysis of the set of equations being considered.

In analysing the nonlinear differential equations entering into
the first group, it is convenient to use the quantities $\rho_t$,
$\rho_\lambda$, $\rho_\varkappa$, $\rho_1$, and $\rho_2$, defined
as follows:
\[
\rho_t(t)=\frac{Y_t(t)}{\tilde{\alpha}_3(t)},\quad
\rho_{\lambda}(t)=\frac{Y_{\lambda}(t)}{\tilde{\alpha}_3(t)},\quad
\rho_{\varkappa}(t)=\frac{Y_{\varkappa}(t)}{\tilde{\alpha}_3(t)},\quad
\rho_1(t)=\frac{\tilde{\alpha}_1(t)}{\tilde{\alpha}_3(t)},\quad
\rho_2(t)=\frac{\tilde{\alpha}_2(t)}{\tilde{\alpha}_3(t)},
\]
where $\tilde{\alpha}_i(t)=g^2_i(t)/(4\pi)^2$,
$Y_t(t)=h^2_t(t)/(4\pi)^2$,
$Y_{\lambda}(t)=\lambda^2(t)/(4\pi)^2$, and
$Y_{\varkappa}(t)=\varkappa^2(t)/(4\pi)^2$.

Let us first consider the simplest case of $\varkappa=0$. The
growth of the Yukawa coupling constant $\lambda(t_0)$ at a fixed
value of $h_t(t_0)$ results in that the Landau pole in solutions
to the renormalization group equations approaches the Grand
Unification scale from above. At a specific value
$\lambda(t_0)=\lambda_{\text{max}}$, perturbation theory at $q\sim
M_X$ cease to be applicable. With increasing (decreasing) Yukawa
coupling constant for the $b$--quark, $\lambda_{\text{max}}$
decreases (increases). In the $(\rho_t,\rho_\lambda)$ plane, the
dependence $\lambda^2_{\text{max}}(h_t^2)$ is represented by a
curve bounding the region of admissible values of the parameters
$\rho_t(t_0)$ and $\rho_\lambda(t_0)$. At $\rho_\lambda=0$, this
curve intersects the abscissa at the point
$\rho_t=\rho_t^{\text{QFP}}(t_0)$. This is the way in which there
arises, in the $(\rho_t,\rho_\lambda)$ plane, the quasi--fixed (or
Hill) line near which solutions to the renormalization group
equations are grouped (see Fig. 1). With increasing $\lambda^2(0)$
and $h_t^2(0)$, the region where the solutions in questions are
concentrated sharply shrinks, and for rather large initial values
of the Yukawa coupling constants they are grouped in a narrow
stripe near the straight line
\begin{equation}
\rho_t(t_0)+0.506\rho_{\lambda}(t_0)=0.91, \label{B7}
\end{equation}
which can be obtained by fitting the results of numerical
calculations (these results are presented in Fig. 2). Moreover,
the combination $h_t^2(t_0)+0.506\lambda^2(t_0)$ of the Yukawa
coupling constants depends much more weakly on $\lambda^2(0)$ and
$h_t^2(0)$ than $\lambda^2(t_0)$ and $h_t^2(t_0)$ individually
\cite{NTB}. In other words, a decrease in $\lambda^2(t_0)$
compensates for an increase in $h_t^2(t_0)$, and vice versa. The
results in Fig. 3, which illustrate the evolution of the above
combinations of the Yukawa coupling constants, also confirm that
this combination is virtually independent of the initial
conditions.

In analysing the results of numerical calculations, our attention
is engaged by a pronounced nonuniformity in the distribution of
solutions to the renormalization group equations along the
infrared quasi--fixed line. The main reason for this is that, in
the regime of strong Yukawa coupling, the solutions in question
are attracted not only to the quasi--fixed but also to the
infrared fixed (or invariant) line. The latter connects two fixed
points. Of these, one is an infrared fixed point of the set of
renormalization group equations within the NMSSM ($\rho_t=7/18$,
$\rho_{\lambda}=0$, $\rho_1=0$, $\rho_2=0$) \cite{B6}, while the
other fixed point $(\rho_\lambda/\rho_t=1)$ corresponds to values
of the Yukawa coupling constants in the region
$Y_t,Y_\lambda\gg\tilde{\alpha}_i$, in which case the gauge
coupling constants on the right--hand sides of the renormalization
group equations can be disregarded \cite{B35}. For the asymptotic
behaviour of the infrared fixed line at $\rho_t,\rho_\lambda\gg 1$
we have
\[
\rho_{\lambda}=\rho_t-\frac{8}{15}-\frac{2}{75}\rho_1,
\]
while in the vicinity of the point $\rho_t=7/18$, $\rho_\lambda=0$
we have
\[
\rho_\lambda\sim(\rho_t-7/18)^{25/14}.
\]

The infrared fixed line is invariant under renormalization group
transformations -- that is, it is independent of the scale at
which the boundary values $Y_t(0)$ and $Y_\lambda(0)$ are
specified and of the boundary values themselves. If the boundary
conditions are such that $Y_t(0)$ and $Y_\lambda(0)$ belong to the
fixed line, the evolution of the Yukawa coupling constants
proceeds further along this line toward the infrared fixed point
of the set of renormalization group equations within the NMSSM.
With increasing $t$, all other solutions to the renormalization
group equations are attracted to the infrared fixed line and, for
$t/(4\pi)\gg 1$, approach the stable infrared fixed point. From
the data in Figs. 1 and 2, it follows that, with increasing
$Y_t(0)$ and $Y_\lambda(0)$, all solutions to the renormalization
group equations are concentrated in the vicinity of the point of
intersection of the infrared fixed and the quasi--fixed line:
\[
\rho^{\text{QFP}}_t(t_0)=0.803,\qquad
\rho^{\text{QFP}}_{\lambda}(t_0)=0.224.
\]
Hence, this point can be considered as the quasi--fixed point of
the set of renormalization group equations within the NMSSM at
$\varkappa=0$.

In a more complicated case where all three Yukawa coupling
constants in the NMSSM are nonzero, analysis of the set of
renormalization group equations presents a much more difficult
problem. In particular, invariant (infrared fixed) and Hill
surfaces come to the fore instead of the infrared fixed and
quasi--fixed lines. For each fixed set of values of the coupling
constants $Y_t(t_0)$ and $Y_\varkappa(t_0)$, an upper limit on
$Y_\lambda(t_0)$ can be obtained from the requirement that
perturbation theory be applicable up to the Grand Unification
scale $M_X$. A change in the values of the Yukawa coupling
constants $h_t$ and $\varkappa$ at the electroweak scale leads to
a growth or a reduction of the upper limit on $Y_\lambda(t_0)$.
The resulting surface in the
$(\rho_t,\rho_\varkappa,\rho_\lambda)$ space is shown in Fig. 4.
In the regime of strong Yukawa coupling, solutions to the
renormalization group equations are concentrated near this
surface. In just the same way as in the case of $Y_\varkappa=0$, a
specific linear combination of $Y_t$, $Y_\lambda$, and
$Y_\varkappa$ is virtually independent of the initial conditions
for $Y_i(0)\to\infty$:
\begin{equation}
\rho_t(t_0)+0.72\rho_{\lambda}(t_0)+0.33\rho_{\varkappa}(t_0)=0.98.
\label{B14}
\end{equation}
The evolution of this combination of Yukawa couplings at various
initial values of the Yukawa coupling constants is illustrated in
Fig. 5.

On the Hill surface, the region that is depicted in Fig. 4 and
near which the solutions in question are grouped shrinks in one
direction with increasing initial values of the Yukawa coupling
constants, with the result that, at $Y_t(0)$, $Y_\varkappa(0)$,
and $Y_\lambda(0)\sim 1$, all solutions are grouped around the
line that appears as the result of intersection of the
quasi--fixed surface and the infrared fixed surface, which
includes the invariant lines lying in the $\rho_\varkappa=0$ and
$\rho_\lambda=0$ planes and connecting the stable infrared point
with, respectively, the fixed point $\rho_\lambda/\rho_t=1$ and
the fixed point $\rho_\varkappa/\rho_t=1$ in the regime of strong
Yukawa coupling. In the limit $\rho_t, \rho_\lambda,
\rho_\varkappa\gg 1$, in which case the gauge coupling constants
can be disregarded, the fixed points
$\rho_\lambda/\rho_t=1,~\rho_\varkappa/\rho_t=0$ and
$\rho_\varkappa\rho_t=1,~\rho_\lambda/\rho_t=0$ cease to be
stable. Instead of them, the stable fixed point
$R_\lambda=3/4,~R_\varkappa=3/8$ \cite{B35} appears in the
$(R_\lambda,R_\varkappa)$ plane, where
$R_\lambda=\rho_\lambda/\rho_t$ and
$R_\varkappa=\rho_\varkappa/\rho_t$. In order to investigate the
behaviour of the solutions to the renormalization group equations
within the NMSSM, it is necessary to linearise the set of these
equations in its vicinity and set $\alpha_i=0$. As a result, we
obtain
\begin{equation}
\begin{split}
R_{\lambda}(t)=&\frac{3}{4}+\left(\frac{1}{2}R_{\lambda 0}+
\frac{1}{\sqrt{5}}R_{\varkappa 0}-\frac{3(\sqrt{5}+1)}{8\sqrt{5}}
\right)\left(\frac{\rho_t(t)}{\rho_{t0}}\right)^{\lambda_1} \\
&{}+\left(\frac{1}{2}R_{\lambda 0}-\frac{1}{\sqrt{5}} R_{\varkappa
0}-\frac{3(\sqrt{5}-1)}{8\sqrt{5}}\right)
\left(\frac{\rho_t(t)}{\rho_{t0}}\right)^{\lambda_2}, \\
R_{\varkappa}(t)=&\frac{3}{8}+\frac{\sqrt{5}}{2}
\left(\frac{1}{2}R_{\lambda 0}+\frac{1}{\sqrt{5}} R_{\varkappa
0}-\frac{3(\sqrt{5}+1)}{8\sqrt{5}}\right)
\left(\frac{\rho_t(t)}{\rho_{t0}}\right)^{\lambda_1} \\
&{}-\frac{\sqrt{5}}{2}\left(\frac{1}{2}R_{\lambda 0}-
\frac{1}{\sqrt{5}}R_{\varkappa 0}-\frac{3(\sqrt{5}-1)}{8\sqrt{5}}
\right)\left(\frac{\rho_t(t)}{\rho_{t0}}\right)^{\lambda_2},\\
\end{split}
\label{B15}
\end{equation}
where $R_{\lambda 0}=R_{\lambda}(0)$, $R_{\varkappa
0}=R_{\varkappa}(0)$, $\rho_{t0}=\rho_t(0)$,
$\lambda_1=\dfrac{3+\sqrt{5}}{9}$,
$\lambda_2=\dfrac{3-\sqrt{5}}{9}$, and
$\rho_t(t)=\dfrac{\rho_{t0}}{1+7\rho_{t0}t}$. From (\ref{B15}), it
follows that the fixed point $R_\lambda=3/4,~R_\varkappa=3/8$
arises as the result of intersection of two fixed lines in the
$(R_\lambda,R_\varkappa)$ plane. The solutions are attracted most
strongly to the line
$\dfrac{1}{2}R_\lambda+\dfrac{1}{\sqrt{5}}R_\varkappa=\dfrac{3}{8}\left(1+\dfrac{1}{\sqrt{5}}\right)$,
since $\lambda_1\gg \lambda_2$. This line passes through three
fixed points in the $(R_\lambda,R_\varkappa)$ plane: $(1,0)$,
$(3/4,3/8)$, and $(0,1)$. In the regime of strong Yukawa coupling,
the fixed line that corresponds, in the
$(\rho_t,\rho_\varkappa,\rho_\lambda)$ space, to the line
mentioned immediately above is that which lies on the invariant
surface containing a stable infrared fixed point. The line of
intersection of the Hill and the invariant surface can be obtained
by mapping this fixed line into the quasi--fixed surface with the
aid of the set of renormalization group equations. For the
boundary conditions, one must than use the values $\lambda^2(0)$,
$\varkappa^2(0)$, and $h_t^2(0)\gg 1$ belonging to the
aforementioned fixed line.

In just the same way as infrared fixed lines, the infrared fixed
surface is invariant under renormalization group transformations.
In the evolution process, solutions to the set of renormalization
group equations within the NMSSM are attracted to this surface. If
boundary conditions are specified n the fixed surface, the ensuing
evolution of the coupling constants proceeds within this surface.
To add further details, we not that, near the surface being
studied and on it, the solutions are attracted to the invariant
line connecting the stable fixed point
$(\rho_\lambda/\rho_t=3/4,~\rho_\varkappa/\rho_t=3/8)$ in the
regime of strong Yukawa coupling with the stable infrared fixed
point within the NMSSM. In the limit
$\rho_t,\rho_\varkappa,\rho_\lambda\gg 1$, the equation for this
line has the form
\begin{equation}
\begin{split}
\rho_{\lambda}&=\frac{3}{4}\rho_t-\frac{176}{417}+
\frac{3}{139}\rho_2-\frac{7} {417}\rho_1, \\
\rho_{\varkappa}&=\frac{3}{8}\rho_t-\frac{56}{417}-
\frac{18}{139}\rho_2-\frac{68}{2085}\rho_1.
\end{split}
\label{B16}
\end{equation}
As one approaches the infrared fixed point, the quantities
$\rho_\lambda$ and $\rho_\varkappa$ tend to zero:
$\rho_\lambda\sim(\rho_t-7/18)^{25/14}$ and
$\rho_\varkappa\sim(\rho_t-7/18)^{9/7}$. This line intersects the
quasi--fixed surface at the point
\[
\rho^{\text{QFP}}_t(t_0)=0.82,\quad
\rho^{\text{QFP}}_{\varkappa}(t_0)=0.087,\quad
\rho^{\text{QFP}}_{\lambda}(t_0)=0.178.
\]
Since all solutions are concentrated in the vicinity of this point
for $Y_t(0), Y_\lambda(0), Y_\varkappa(0)\to\infty$, it should be
considered as a quasi--fixed point for the set of renormalization
group equations within the NMSSM. We note, however, that the
solutions are attracted to the invariant line (\ref{B16}) and to
the quasi--fixed line on the Hill surface. This conclusion can be
drawn from the an analysis of the behaviour of the solutions near
the fixed point $(R_\lambda=3/4,~R_\varkappa=3/8)$ (see
(\ref{B15})). Once the solutions have approached the invariant
line
$\dfrac{1}{2}R_\lambda+\dfrac{1}{\sqrt{5}}R_\varkappa=\dfrac{3}{8}\left(1+\dfrac{1}{\sqrt{5}}\right)$,
their evolution is governed by the expression
$(\epsilon(t))^{0.085}$, where $\epsilon(t)=\rho_t(t)/\rho_{t0}$.
This means that the solutions begin to be attracted to the
quasi--fixed point and to the invariant line (\ref{B16}) with a
sizable strength only when $Y_i(0)$ reaches a value of $10^2$, at
which perturbation theory is obviously inapplicable. Thus, it is
not the infrared quasi--fixed point but the quasi--fixed line on
the Hill surface (see Fig. 4) that, within the NMSSM, plays a key
role in analysing the behaviour of the solutions to the
renormalization group equations in the regime of strong Yukawa
coupling, where all $Y_i(0)$ are much greater than
$\tilde{\alpha}_0$.

\section{Renormalization of the soft SUSY breaking parameters}

If the evolution of gauge and Yukawa coupling constants is known,
the remaining subset of renormalization group equations within the
MNSSM can be treated as a set of linear differential equations for
the parameters of a soft breakdown of supersymmetry. For universal
boundary conditions, a general solution for the trilinear coupling
constants $A_i(t)$ and for the masses of scalar fields $m_i^2(t)$
has the form
\begin{gather}
A_i(t)=e_i(t)A+f_i(t)M_{1/2}\, , \label{C8} \\
m_i^2(t)=a_i(t)m_0^2+b_i(t)M_{1/2}^2+c_i(t)AM_{1/2}+d_i(t)A^2\, .
\label{C9}
\end{gather}
The functions $e_i(t)$, $f_i(t)$, $a_i(t)$, $b_i(t)$, $c_i(t)$,
and $d_i(t)$, which determine the evolution of $A_i(t)$ and
$m_i^2(t)$, remain unknown, since an analytic solution to the full
set of renormalization group equations within the NMSSM is
unavailable. These functions greatly depend on the choice of
values for the Yukawa coupling constants at the Grand Unification
scale $M_X$. At the electroweak scale $t=t_0$, relations
(\ref{C8}) and (\ref{C9}) specify the parameters $A_i^2(t_0)$ and
$m_i^2(t_0)$ of a soft breaking of supersymmetry as functions of
their initial values at the Grand Unification scale.

The results of our numerical analysis indicate that, with
increasing $Y_i(0)$, where $Y_t(t)=\dfrac{h_t^2(t)}{(4\pi)^2}$,
$Y_\lambda(t)=\dfrac{\lambda^2(t)}{(4\pi)^2}$, and
$Y_\varkappa(t)=\dfrac{\varkappa^2(t)}{(4\pi)^2}$, the functions
$e_i(t_0)$, $c_i(t_0)$, and $d_i(t_0)$ decrease and tend to zero
in the limit $Y_i(0)\to\infty$, relations (\ref{C8}) and
(\ref{C9}) becoming much simpler in this limit. Instead of the
squares of the scalar particle masses, it is convenient to
consider their linear combinations
\begin{equation}
\begin{split}
\mathfrak{M}_t^2(t)&=m_2^2(t)+m_Q^2(t)+m_U^2(t),\\
\mathfrak{M}_{\lambda}^2(t)&=m_1^2(t)+m_2^2(t)+m_y^2(t),\\
\mathfrak{M}_{\varkappa}^2(t)&=3m_y^2(t)
\end{split}
\label{C10}
\end{equation}
in analysing the set of renormalization group equations. In the
case of universal boundary conditions, the solutions to the
differential equations for $\mathfrak{M}_i^2(t)$ can be
represented in the same form as the solutions for $m_i^2(t)$ (see
(\ref{C9})); that is
\begin{equation}
\mathfrak{M}_i^2(t)=3\tilde{a}_i(t)m_0^2+\tilde{b}_i(t)M_{1/2}^2+
\tilde{c}_i(t)A M_{1/2}+\tilde{d}_i(t)A^2. \label{C11}
\end{equation}
Since the homogeneous equations for $A_i(t)$ and
$\mathfrak{M}_i^2(t)$ have the same form, the functions
$\tilde{a}_i(t)$ and $e_i(t)$ coincide; in the limit of strong
Yukawa coupling, the $m_0^2$ dependence disappears in the
combinations (\ref{C10}) of the scalar particle masses as the
solutions to the renormalization group equations for the Yukawa
coupling constants approach quasi--fixed points. This behaviour of
the solutions implies that $A_i(t)$ and $\mathfrak{M}_i^2(t)$
corresponding to $Y_i(0)\gg\tilde{\alpha}_i(0)$ also approach
quasi--fixed points. As we see in the previous section, two
quasi--fixed points of the renormalization group equations within
the NMSSM are of greatest interest from the physical point of
view. Of these, one corresponds to the boundary conditions
$Y_t(0)=Y_\lambda(0)\gg\tilde{\alpha}_i(0)$ and $Y_\varkappa(0)=0$
for the Yukawa coupling constants. The fixed points calculated for
the parameters of a soft breaking of supersymmetry by using these
values of the Yukawa coupling constants are
\begin{equation}
\begin{aligned} \rho_{A_t}^{\text{QFP}}(t_0)&\approx 1.77,
&\rho_{\mathfrak{M}^2_t}^{\text{QFP}}(t_0)&\approx 6.09,\\
\rho_{A_{\lambda}}^{\text{QFP}}(t_0)&\approx -0.42,\qquad
&\rho_{\mathfrak{M}^2_{\lambda}}^{\text{QFP}}(t_0)&\approx -2.28,
\end{aligned}
\label{C12}
\end{equation}
where $\rho_{A_i}(t)=A_i(t)/M_{1/2}$ and
$\rho_{\mathfrak{M}_i^2}(t)=\mathfrak{M}_i^2/M_{1/2}^2$. Since the
coupling constant $\varkappa$ for the self--interaction of neutral
scalar fields is small in the case being considered,
$A_\varkappa(t)$ and $\mathfrak{M}_\varkappa^2(t)$ do not approach
the quasi--fixed point. Nonetheless, the spectrum of SUSY
particles is virtually independent of the trilinear coupling
constant $A_\varkappa$ since $\varkappa\to 0$.

In just the same way, one can determine the position of the other
quasi--fixed point for $A_i(t)$ and $\mathfrak{M}_i^2(t)$, that
which corresponds to $R_{\lambda 0}=3/4,~R_{\varkappa 0}=3/8$. The
results are
\begin{equation}
\begin{aligned} \rho_{A_t}^{\text{QFP}}(t_0)&\approx 1.73,
&\rho_{A_{\lambda}}^{\text{QFP}}(t_0)&\approx -0.43,
&\rho_{A_{\varkappa}}^{\text{QFP}}(t_0)&\approx 0.033,\\
\rho_{\mathfrak{M}^2_t}^{\text{QFP}}(t_0)&\approx 6.02,\quad
&\rho_{\mathfrak{M}^2_{\lambda}}^{\text{QFP}}(t_0)&\approx
-2.34,\quad
&\rho_{\mathfrak{M}^2_{\varkappa}}^{\text{QFP}}(t_0)&\approx 0.29,
\end{aligned}
\label{C13}
\end{equation}
where $R_{\lambda 0}=Y_\lambda(0)/Y_t(0)$ and $R_{\varkappa
0}=Y_\varkappa(0)/Y_t(0)$. It should be noted that, in the
vicinities of quasi--fixed points, we have
$\rho_{\mathfrak{M}^2_{\lambda}}^{\text{QFP}}(t_0)<0$. Negative
values of $\mathfrak{M}_\lambda^2(t_0)$ lead to a negative value
of the parameter $m_2^2(t_0)$ in the potential of interaction of
Higgs fields. In other words, an elegant mechanism that is
responsible for a radiative violation of $SU(2)\otimes U(1)$
symmetry and which does not require introducing tachyons in the
spectrum of the theory from the outset survives in the regime of
strong Yukawa coupling within the NMSSM. This mechanism of gauge
symmetry breaking was first discussed in \cite{C30} by considering
the example of the minimal SUSY model.

By using the fact that $\mathfrak{M}_i^2(t)$ as determined for the
case of universal boundary conditions is virtually independent of
$m_0^2$, we can predict $a_i(t_0)$ values near the quasi--fixed
points (see \cite{NTC}). The results are
\begin{equation}
\begin{split}
1)~&R_{\lambda 0}=1,~R_{\varkappa 0}=0,\\
&a_y(t_0)=a_u(t_0)=\frac{1}{7},~a_1(t_0)=a_q(t_0)=\frac{4}{7},~
a_2(t_0)=-\frac{5}{7}\, ;\\ 2)~&R_{\lambda 0}=3/4,~R_{\varkappa
0}=3/8,\\ &a_y(t_0)=0,~
a_1(t_0)=-a_2(t_0)=\frac{2}{3},~a_q(t_0)=\frac{5}{9},~
a_u(t_0)=\frac{1}{9}\, .
\end{split}
\label{C14}
\end{equation}
To do this, it was necessary to consider specific combinations of
the scalar particle masses, such as $m_U^2-2m_Q^2$,
$m_Q^2+m_U^2-m_2^2+m_1^2$, and $m_y^2-2m_1^2$ (at $\varkappa=0$),
that are not renormalized by Yukawa interactions. As a result, the
dependence of the above combinations of the scalar particle masses
on $m_0^2$ at the electroweak scale is identical to that at the
Grand Unification scale. The predictions in (\ref{C14}) agree
fairly well with the results of numerical calculations.

Let us now consider the case of nonuniversal boundary conditions
for the soft SUSY breaking parameters. The results of our
numerical analysis, which are illustrated in Figs. 6 and 7,
indicate that, in the vicinity of the infrared fixed point at
$Y_\varkappa=0$, solutions to the renormalization group equations
at the electroweak scale are concentrated near some straight lines
for the case where the simulation was performed by using boundary
conditions uniformly distributed in the $(A_t,A_\lambda)$ and the
$(\mathfrak{M}_t^2,\mathfrak{M}_\lambda^2)$ plane. The strength
with which these solutions are attracted to them grows with
increasing $Y_i(0)$. The equations for the lines being considered
can be obtained by fitting the numerical results displayed in
Figs. 6 and 7. This yields
\begin{equation}
\begin{gathered}
A_t+0.147 A_{\lambda}=1.70M_{1/2},\\ \mathfrak{M}^2_t+0.147
\mathfrak{M}^2_{\lambda}=5.76 M_{1/2}^2 .
\end{gathered}
\label{C18}
\end{equation}
For $Y_\varkappa(0)\gg\tilde{\alpha}_0$ solutions to the
renormalization group equations are grouped near planes in the
space of the parameters of a soft breaking of supersymmetry
$(A_t,A_\lambda,A_\varkappa)$ and
$(\mathfrak{M}_t^2,\mathfrak{M}_\lambda^2,\mathfrak{M}_\varkappa^2)$
(see Figs. 8-10):
\begin{equation}
\begin{gathered}
A_t+0.128 A_{\lambda}+0.022 A_{\varkappa}=1.68 M_{1/2} ,\\
\mathfrak{M}^2_t+0.128 \mathfrak{M}^2_{\lambda}+0.022
\mathfrak{M}^2_{\varkappa}=5.77 M_{1/2}^2 .
\end{gathered}
\label{C19}
\end{equation}
It can be seen from Figs. 8 and 9 that, as the values of the
Yukawa coupling constants at the Grand Unification scale are
increased, the areas of the surfaces near which the solutions
$A_i(t)$ and $\mathfrak{M}_i^2(t)$ are concentrated shrink in one
of the directions, with the result that, at $Y_i(0)\sim 1$, the
solutions to the renormalization group equations are attracted to
one of the straight lines belonging to these surfaces.

The numerical calculations also showed that, with increasing
$Y_i(0)$, only in the regime of infrared quasi--fixed points (that
is, at $R_{\lambda 0}=1,~R_{\varkappa 0}=0$ or at $R_{\lambda
0}=3/4,~R_{\varkappa 0}=3/8$) $e_i(t_0)$ and $\tilde{a}_i(t_0)$
decrease quite fast, in proportion to $1/Y_i(0)$. Otherwise, the
dependence on $A$ and $m_0^2$ disappears much more slowly with
increasing values of the Yukawa coupling constants at the Grand
Unification scale -- specifically, in proportion to
$(Y_i(0))^{-\delta}$, where $\delta<1$ (for example,
$\delta=0.35-0.40$ at $\varkappa=0$). In the case of nonuniversal
boundary conditions, only when solutions to the renormalization
group equations approach quasi--fixed points are these solutions
attracted to the fixed lines and surfaces in the space of the
parameters of a soft breaking of supersymmetry, and in the limit
$Y_i(0)\to\infty$, the parameters $A_i(t)$ and $\mathfrak{M}_i
^2(t)$ cease to be dependent on the boundary conditions.

For the solutions of the renormalization group equations for the
soft SUSY breaking parameters near the electroweak scale in the
strong Yukawa coupling regime one can construct an expansion in
powers of the small parameter $\epsilon_t(t)=Y_t(t)/Y_t(0)$:
\begin{equation}
\binom{A_i(t)}{\mathfrak{M}^2_i(t)}=\sum_k u_{ik}v_{ik}(t)
\binom{\alpha_k}{\beta_k}(\epsilon_t(t))^{\lambda_k}+\dots,
\label{C25}
\end{equation}
where $\alpha_i$ and $\beta_i$ are constants of integration that
can be expressed in terms of $A_i(0)$ and $\mathfrak{M}^2_i(0)$.
The functions $v_{ij}(t)$ are weakly dependent on the Yukawa
coupling constants at the scale $M_X$, and $v_{ij}(0)=1$. They
appear upon renormalizing the parameters of a soft breaking of
supersymmetry from $q\sim 10^{12}-10^{13}\text{~GeV}$ to $q\sim
m_t$. In equations (\ref{C25}), we have omitted terms proportional
to $M_{1/2}$, $M_{1/2}^2$, $A_i(0)M_{1/2}$, and $A_i(0)A_j(0)$.

At $\varkappa=0$, we have two eigenvalues and two corresponding
eigenvectors:
\[
\lambda=\begin{pmatrix} 1 \\ 3/7 \end{pmatrix},\qquad
u=\begin{pmatrix} 1 & 1
\\ 1 & -3
\end{pmatrix},
\]
whose components specify $(A_t,A_\lambda)$ and
$(\mathfrak{M}_t^2,\mathfrak{M}_\lambda^2)$. With increasing
$Y_t(0)\simeq Y_\lambda(0)$, the dependence on $\alpha_0$ and
$\beta_0$ becomes weaker and the solutions at $t=t_0$ are
concentrated near the straight lines
$(A_t(\alpha_1),A_\lambda(\alpha_1))$ and
$(\mathfrak{M}_t^2(\beta_1),\mathfrak{M}_\lambda^2(\beta_1))$. In
order to obtain the equations for these straight lines, it is
necessary to set $A_\lambda(0)=-3A_t(0)$ and
$\mathfrak{M}_\lambda^2(0)=-3\mathfrak{M}_t^2(0)$ at the Grand
Unification scale. At the electroweak scale, there then arise a
relation between $A_t(t_0)$ and $A_\lambda(t_0)$ and a relation
between $\mathfrak{M}_t^2(t_0)$ and $\mathfrak{M}_\lambda^2(t_0)$:
\begin{equation}
\begin{gathered}
A_t+0.137 A_{\lambda}=1.70 M_{1/2},\\ \mathfrak{M}^2_t+ 0.137
\mathfrak{M}^2_{\lambda}=5.76 M^2_{1/2}.
\end{gathered}
\label{C26}
\end{equation}
These relations agree well with the equations deduced for the
straight lines at $Y_i(0)\sim 1$ by fitting the results of the
numerical calculations (\ref{C18}).

When the Yukawa coupling constant $\varkappa$ is nonzero, we have
three eigenvalues and three corresponding eigenvectors:
\[
\lambda=\begin{pmatrix} 1 \\ \frac{3+\sqrt{5}}{9} \\
\frac{3-\sqrt{5}}{9}
\end{pmatrix},\qquad
u=\begin{pmatrix} 1 & -\frac{1+\sqrt{5}}{24} &
\frac{\sqrt{5}-1}{24} \\ 1 & \frac{\sqrt{5}}{6} &
-\frac{\sqrt{5}}{6} \\ 1 & 1 & 1
\end{pmatrix},
\]
whose components specify $(A_t,A_\lambda,A_\varkappa)$ and
$(\mathfrak{M}_t^2,\mathfrak{M}_\lambda^2,\mathfrak{M}_\varkappa^2)$.
An increase in $Y_\lambda(0)\simeq 2Y_\varkappa(0)\simeq
\dfrac{3}{4}Y_t(0)$ leads to the following: first, the dependence
of $A_i(t)$ and $\mathfrak{M}_i^2(t)$ on $\alpha_0$ and $\beta_0$
disappears, which leads to the emergence of planes in the space
spanned by the parameters of a soft breaking of supersymmetry:
\begin{equation}
\begin{gathered}
A_t+0.103 A_{\lambda}+0.0124 A_{\varkappa}=1.69 M_{1/2},\\
\mathfrak{M}^2_t+0.103 \mathfrak{M}^2_{\lambda}+0.0124
\mathfrak{M}^2_{\varkappa}=5.78 M^2_{1/2}.
\end{gathered}
\label{C28}
\end{equation}
After that, the dependence on $\alpha_1$ and $\beta_1$ becomes
weaker at $Y_i(0)\sim 1$. This means that, with increasing initial
values of the Yukawa coupling constants, solutions to the
renormalization group equations are grouped near some straight
lines and we can indeed see precisely this pattern in Figs. 8-10.
All equations presented here for the straight lines and planes in
the $\mathfrak{M}_i^2$ space were obtained at $A_i(0)=0$.

From relations (\ref{C26}) and (\ref{C28}), it follows that
$A_t(t_0)$ and $\mathfrak{M}_t^2(t_0)$ are virtually independent
of the initial conditions; that is, the straight lines and planes
are orthogonal to the $A_t$ and $\mathfrak{M}_t^2$ axes. On the
other hand, the $A_\varkappa(t_0)$ and
$\mathfrak{M}_\varkappa^2(t_0)$ values that correspond to the
Yukawa self--interaction constant $Y_\varkappa$ for the neutral
fields are fully determined by the boundary conditions for the
parameters of a soft breaking of supersymmetry. For this reason,
the planes in the $(A_t,A_\lambda,A_\varkappa)$ and
$(\mathfrak{M}_t^2,\mathfrak{M}_\lambda^2,\mathfrak{M}_\varkappa^2)$
spaces are virtually parallel to the $A_\varkappa$ and
$\mathfrak{M}_\varkappa^2$ axes.

\section{Conclusions}

In the strong Yukawa coupling regime in the NMSSM, solutions to
the renormalisation group equations for $Y_i(t)$ are attracted to
quasi--fixed lines and surfaces in the space of Yukawa coupling
constants and specific combinations of $\rho_i(t)$ are virtually
independent of their initial values at the Grand Unification
scale. For $Y_i(0)\to\infty$, all solutions to the renormalisation
group equations are concentrated near quasi--fixed points. These
points emerge as the result of intersection of Hill lines or
surfaces with the invariant line that connects the stable fixed
point for $Y_i\gg\tilde{\alpha}_i$ with the stable infrared fixed
point. For the renormalisation group equations within the NMSSM,
we have listed all the most important invariant lines and surfaces
and studied their asymptotic behaviour for
$Y_i\gg\tilde{\alpha}_i$ and in the vicinity of the infrared fixed
point.

With increasing $Y_i(0)$, the solutions in question approach
quasi--fixed points quite slowly; that is, the deviation is
proportional to $(\epsilon_t(t))^\delta$, where
$\epsilon_t(t)=Y_t(t)/Y_t(0)$ and $\delta$ is calculated by
analysing the set of the renormalisation group equations in the
regime of strong Yukawa coupling. As a rule, $\delta$ is positive
and much less than unity. By way of example, we indicate that, in
the case where all three Yukawa coupling constants differ from
zero, $\delta\approx 0.085$. Of greatest importance in analysing
the behaviour of solutions to the renormalisation group equations
within the NMSSM at $Y_t(0),Y_\lambda(0),Y_\varkappa(0)\sim 1$ is
therefore not the infrared quasi--fixed point but the line lying
on the Hill surface and emerging as the intersection of the Hill
and invariant surface. This line can be obtained by mapping the
fixed points $(1,0)$, $(3/4,3/8)$, and $(0,1)$ in the
$(R_\lambda,R_\varkappa)$ plane for $Y_i\gg\tilde{\alpha}_i$ into
the quasi--fixed surface by means of renormalisation group
equations.

While $Y_i(t)$ approach quasi--fixed points, the corresponding
solutions for the trilinear coupling constants $A_i(t)$
characterising scalar fields and for the combinations
$\mathfrak{M}_i^2(t)$ of the scalar particle masses (see
(\ref{C10})) cease to be dependent on their initial values at the
scale $M_X$ and, in the limit $Y_i(0)\to\infty$, also approach the
fixed points in the space spanned by the parameters of a soft
breaking of supersymmetry. Since the set of differential equations
for $A_i(t)$ and $m_i^2(t)$ is linear, the $A$, $M_{1/2}$, and
$m_0^2$ dependence of the parameters of a soft breaking of
supersymmetry at the electroweak scale can be explicitly obtained
for universal boundary conditions. It turns out that, near the
quasi--fixed points, all $A_i(t)$ and all $\mathfrak{M}_i^2(t)$
are proportional to $M_{1/2}$ and $M_{1/2}^2$, respectively. Thus,
we have shown that, in the parameter space region considered here,
the solutions to the renormalization group equations for the
trilinear coupling constants and for some combinations of the
scalar particle masses are focused in a narrow interval within the
infrared region. Since the neutral scalar field $Y$ is not
renormalized by gauge interactions, $A_\varkappa(t)$ and
$\mathfrak{M}_\varkappa^2(t)$ are concentrated near zero;
therefore they are still dependent on the initial conditions. The
parameters $A_t(t_0)$ and $\mathfrak{M}_t^2(t_0)$ show the weakest
dependence on $A$ and $m_0^2$ because these parameters are
renormalized by strong interactions. By considering that the
quantities $\mathfrak{M}_i^2(t_0)$ are virtually independent of
the boundary conditions, we have calculated, near the quasi--fixed
points, the values of the scalar particle masses at the
electroweak scale.

In the general case of nonuniversal boundary conditions, the
solutions to the renormalization group equations within the NMSSM
for $A_i(t)$ and $\mathfrak{M}_i^2(t)$ are grouped near some
straight lines and planes in the space spanned by the parameters
of a soft breaking of supersymmetry. Moving along these lines and
surfaces as $Y_i(0)$ is increased, the trilinear coupling
constants and the above combinations of the scalar particle masses
approach quasi--fixed points. However, the dependence of these
couplings on $A_i(0)$ and $\mathfrak{M}_i^2(0)$ dies out quite
slowly, in proportion to $(\epsilon_t(t))^{\lambda}$, where
$\lambda$ is a small positive number; as a rule, $\lambda\ll 1$.
For example, $\lambda=3/7$ at $Y_\varkappa=0$ and $\lambda\approx
0.0085$ at $Y_\varkappa\ne 0$. The above is invalid only for the
solutions $A_i(t)$ and $\mathfrak{M}_i^2(t)$ that correspond to
universal boundary conditions for the parameters of a soft
breaking of supersymmetry and to the initial values of $R_{\lambda
0}=1,~R_{\varkappa 0}=0$ and $R_{\lambda 0}=3/4,~R_{\varkappa
0}=3/8$ for the Yukawa coupling constants at the Grand Unification
scale. They correspond to quasi--fixed points of the
renormalization group equations for $Y_i(t)$. As the Yukawa
coupling constants are increased, such solutions are attracted to
infrared quasi--fixed points in proportion to $\epsilon_t(t)$.

Straight lines in the $(A_t,A_\lambda,A_\varkappa)$ and
$(\mathfrak{M}_t^2,\mathfrak{M}_\lambda^2,\mathfrak{M}_\varkappa^2)$
spaces play a key role in the analysis of the behaviour of
solutions for $A_i(t)$ and $\mathfrak{M}_i^2(t)$ in the case where
$Y_t(0),Y_\lambda(0),Y_\varkappa(0)\sim 1$. In the space spanned
by the parameters of a soft breaking of supersymmetry, these
straight lines lie in the planes near which $A_i(t)$ and
$\mathfrak{M}_i^2(t)$ are grouped in the regime of strong Yukawa
coupling at the electroweak scale. The straight lines and planes
that were obtained by fitting the results of numerical
calculations are nearly orthogonal to the $A_t$ and
$\mathfrak{M}_t^2$ axes. This is because the constants $A_t(t_0)$
and $\mathfrak{M}_t^2(t_0)$ are virtually independent of the
initial conditions at the scale $M_X$. On the other hand, the
parameters $A_\varkappa(t_0)$ and $\mathfrak{M}_\varkappa^2(t_0)$
are determined, to a considerable extent, by the boundary
conditions at the scale $M_X$. At $R_{\lambda 0}=3/4$ and
$R_{\varkappa 0}=3/8$, the planes in the
$(A_t,A_\lambda,A_\varkappa)$ and
$(\mathfrak{M}_t^2,\mathfrak{M}_\lambda^2,\mathfrak{M}_\varkappa^2)$
spaces are therefore parallel to the $A_\varkappa$ and
$\mathfrak{M}_\varkappa^2$ axes.

\section*{Acknowledgements}

The authors are grateful to M. I. Vysotsky, D. I. Kazakov, L. B.
Okun, and K. A. Ter--Martirosyan for stimulating questions,
enlightening discussions and comments. R. B. Nevzorov is indebted
to DESY Theory Group for hospitality extended to him.

This work was supported by the Russian Foundation for Basic
Research (RFBR), projects \#\# 00-15-96786, 00-15-96562, and
02-02-17379.

\newpage

\section*{Figure captions}

{\bfseries Fig.~1.} The values of the Yukawa couplings at the
electroweak scale corresponding to the initial values at the GUT
scale uniformly distributed in a square $2\le
h_t^2(0),\lambda^2(0)\le 10$. The thick and thin curves represent,
respectively, the invariant and the Hill line. The dashed line is
a fit of the values $(\rho_t(t_0),\rho_\lambda(t_0))$ for $20\le
h_t^2(0),\lambda^2(0)\le 100$.\\

{\bfseries Fig.~2.} The values of the Yukawa couplings at the
electroweak scale corresponding to the initial values at the GUT
scale uniformly distributed in a square $20\le
h_t^2(0),\lambda^2(0)\le 100$. The dashed line is a fit of the
values $(\rho_t(t_0),\rho_\lambda(t_0))$ for $20\le
h_t^2(0),\lambda^2(0)\le 100$.\\

{\bfseries Fig.~3.} Evolution of the combination
$\rho_t(t)+0.506\rho_\lambda(t)$ of the Yukawa couplings from the
GUT scale ($t=0$) to the electroweak scale ($t=t_0$) for
$\varkappa^2=0$ and for various initial values $h_t^2(0)$ -- Fig.
3a, $\lambda^2(0)$ -- Fig. 3b.\\

{\bfseries Fig.~4.} Quasi--fixed surface in the
$(\rho_t,\rho_\varkappa,\rho_\lambda)$ space. The shaded part of
the surface represents the region near which the solutions
corresponding to the initial values $2\le
h_t^2(0),\varkappa^2(0),\lambda^2(0)\le 10$ -- Fig. 4a, $20\le
h_t^2(0),\varkappa^2(0),\lambda^2(0)\le 100$ -- Fig. 4b are
concentrated.\\

{\bfseries Fig.~5.} Evolution of the combination
$\rho_t+0.720\rho_\lambda+0.3330\rho_\varkappa$ of the Yukawa
couplings from the GUT scale ($t=0$) to the electroweak scale
($t=t_0$) for various initial values $h_t^2(0)$ -- Fig. 5a,
$\lambda^2(0)$ -- Fig. 5b, $\varkappa^2(0)$ -- Fig. 5c.\\

{\bfseries Fig.~6.} The values of the trilinear couplings $A_t$
and $A_\lambda$ at the electroweak scale corresponding to the
initial values uniformly distributed in the $(A_t,A_\lambda)$
plane, calculated at $\varkappa^2=0$ and
$h_t^2(0)=\lambda^2(0)=20$. The straight line is a fit of the
values $(A_t(t_0),A_\lambda(t_0))$.\\

{\bfseries Fig.~7.} The values of the combinations of masses
$\mathfrak{M}_t^2$ and $\mathfrak{M}_\lambda^2$ at the electroweak
scale corresponding to the initial values uniformly distributed in
the
$(\mathfrak{M}_t^2/M_{1/2}^2,\mathfrak{M}_\lambda^2/M_{1/2}^2)$
plane, calculated at $\varkappa^2=0$, $h_t^2(0)=\lambda^2(0)=20$,
and $A_t(0)=A_\lambda(0)=0$. The straight line is a fit of the
values $(\mathfrak{M}_t^2(t_0),\mathfrak{M}_\lambda^2(t_0))$.\\

{\bfseries Fig.~8.} Planes in the parameter spaces
$(A_t/M_{1/2},A_\lambda/M_{1/2},A_\varkappa/M_{1/2})$ -- Fig. 8a,
and
$(\mathfrak{M}_t^2/M_{1/2}^2,\mathfrak{M}_\lambda^2/M_{1/2}^2,\mathfrak{M}_\varkappa^2/M_{1/2}^2)$
-- Fig. 8b. The shaded parts of the planes correspond to the
regions near which the solutions at $h_t^2(0)=16$,
$\lambda^2(0)=12$, and $\varkappa^2(0)=6$ are concentrated. The
initial values $A_i(0)$ and $\mathfrak{M}_i^2(0)$ vary in the
ranges $-M_{1/2}\le A\le M_{1/2}$ and $0\le\mathfrak{M}_i^2(0)\le
3M_{1/2}^2$, respectively.\\

{\bfseries Fig.~9.} Planes in the parameter spaces
$(A_t/M_{1/2},A_\lambda/M_{1/2},A_\varkappa/M_{1/2})$ -- Fig. 9a,
and
$(\mathfrak{M}_t^2/M_{1/2}^2,\mathfrak{M}_\lambda^2/M_{1/2}^2,\mathfrak{M}_\varkappa^2/M_{1/2}^2)$
-- Fig. 9b. The shaded parts of the planes correspond to the
regions near which the solutions at $h_t^2(0)=32$,
$\lambda^2(0)=24$, and $\varkappa^2(0)=12$ are concentrated. The
initial values $A_i(0)$ and $\mathfrak{M}_i^2(0)$ vary in the
ranges $-M_{1/2}\le A\le M_{1/2}$ and $0\le\mathfrak{M}_i^2(0)\le
3M_{1/2}^2$, respectively.\\

{\bfseries Fig.~10.} Set of points in planes
$(0.0223(A_{\varkappa}/M_{1/2})+0.1278(A_{\lambda}/M_{1/2}),~A_t/M_{1/2})$
-- Fig. 10a, and
$(0.0223(\mathfrak{M}^2_{\varkappa}/M_{1/2}^2)+0.1278(\mathfrak{M}^2_{\lambda}/M_{1/2}^2),
~\mathfrak{M}^2_t/M^2_{1/2})$ -- Fig. 10b, corresponding to the
values of parameters of soft SUSY breaking for $h_t^2(0)=32$,
$\lambda^2(0)=24$, $\varkappa^2(0)=12$, and for a uniform
distribution of the boundary conditions in the parameter spaces
$(A_t,A_\lambda,A_\varkappa)$ and $(\mathfrak{M}^2_t,
\mathfrak{M}^2_{\lambda},\mathfrak{M}^2_{\varkappa})$. The initial
values $A_i(0)$ and $\mathfrak{M}_i^2(0)$ vary in the ranges
$-M_{1/2}\le A\le M_{1/2}$ and $0\le\mathfrak{M}_i^2(0)\le
3M_{1/2}^2$, respectively. The straight lines in Figs. 10a and 10b
correspond to the planes in Figs. 9a and 9b, respectively.

\newpage

\begin{center}

\includegraphics[height=100mm,keepaspectratio=true]{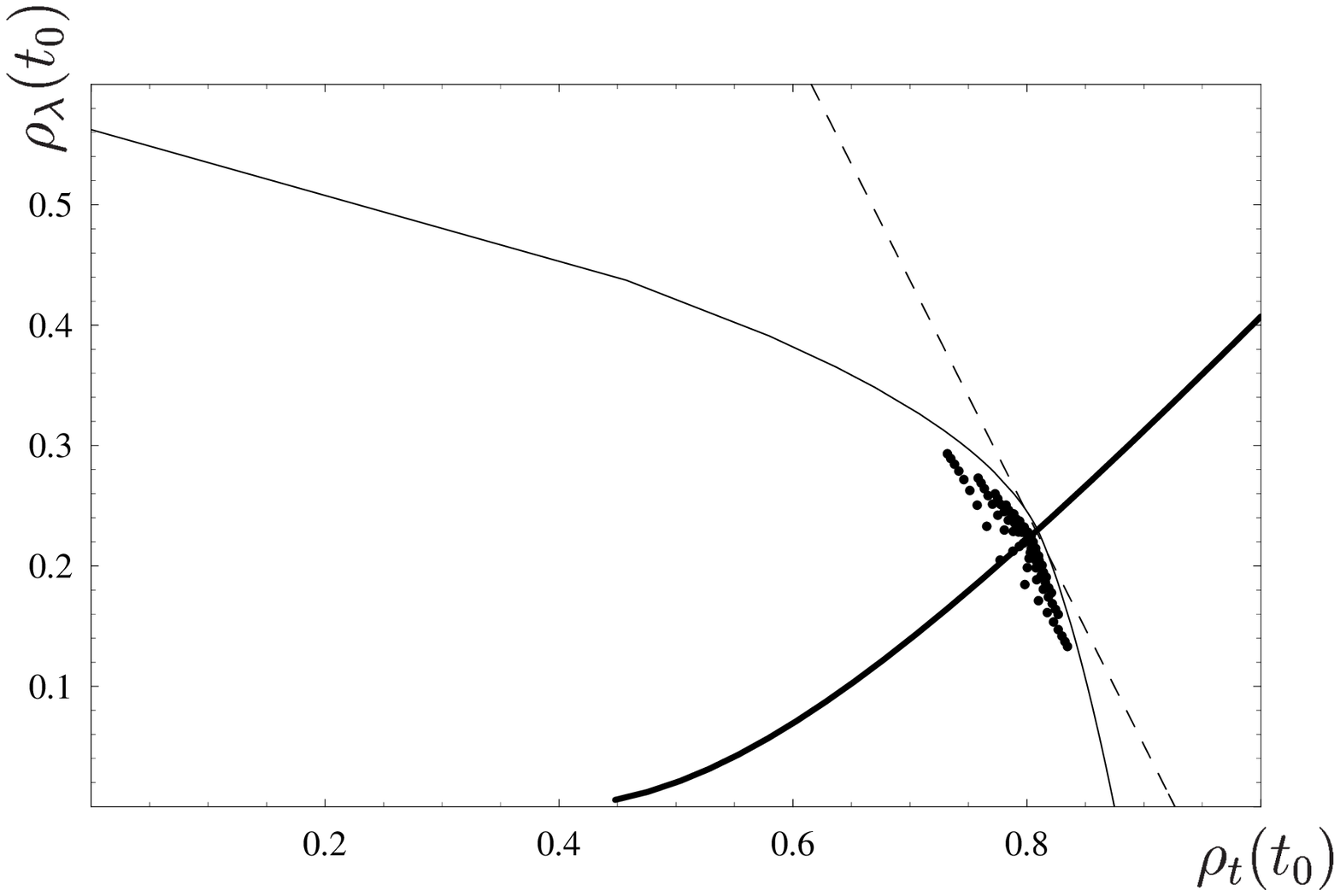}

\vspace{2mm}\hspace*{2mm}{\large\bfseries Fig.1.}

\vspace{15mm}\includegraphics[height=100mm,keepaspectratio=true]{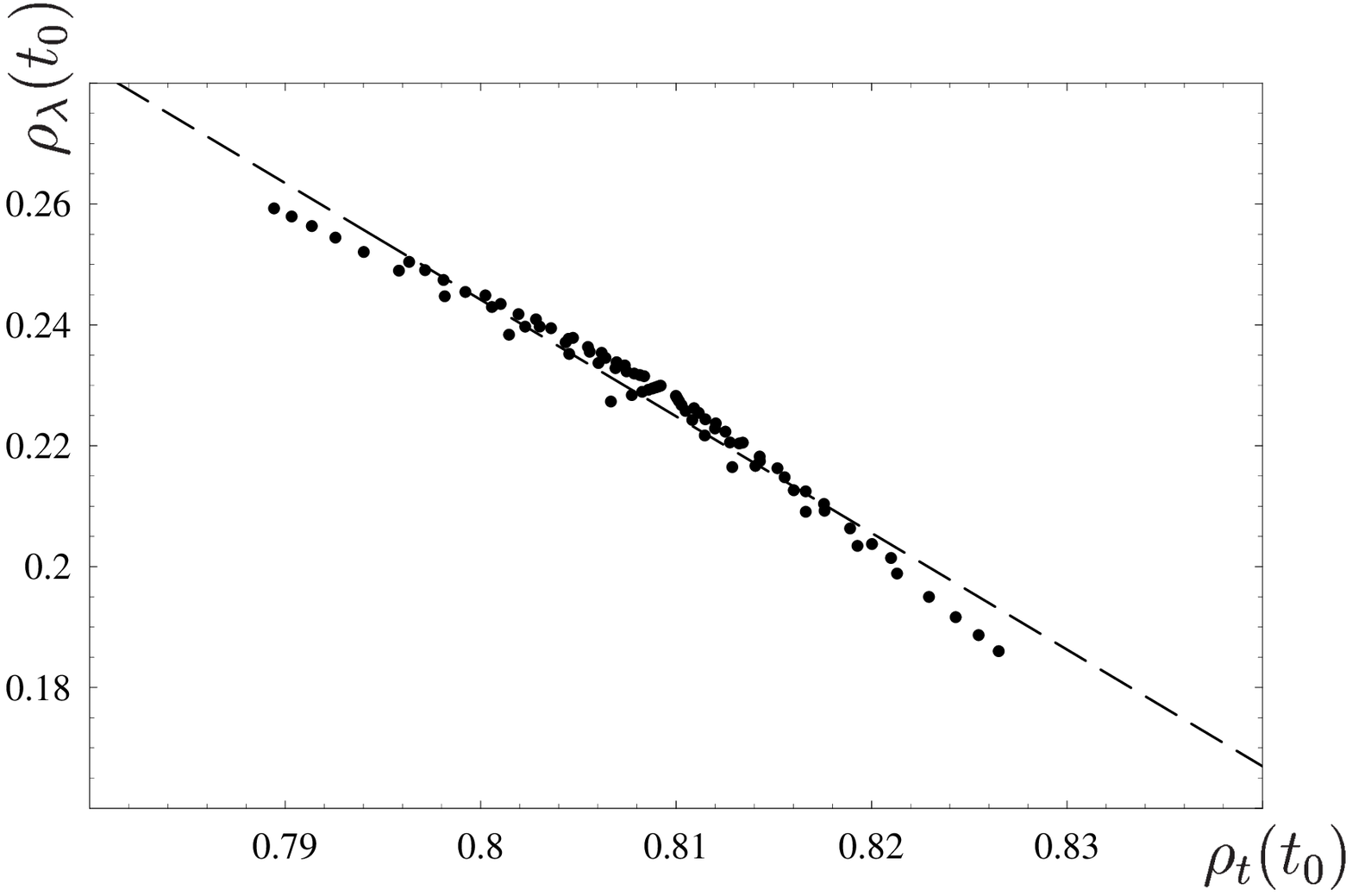}

\vspace{2mm}\hspace*{2mm}{\large\bfseries Fig.2.}

\end{center}

\newpage

\begin{center}

\includegraphics[height=96mm,keepaspectratio=true]{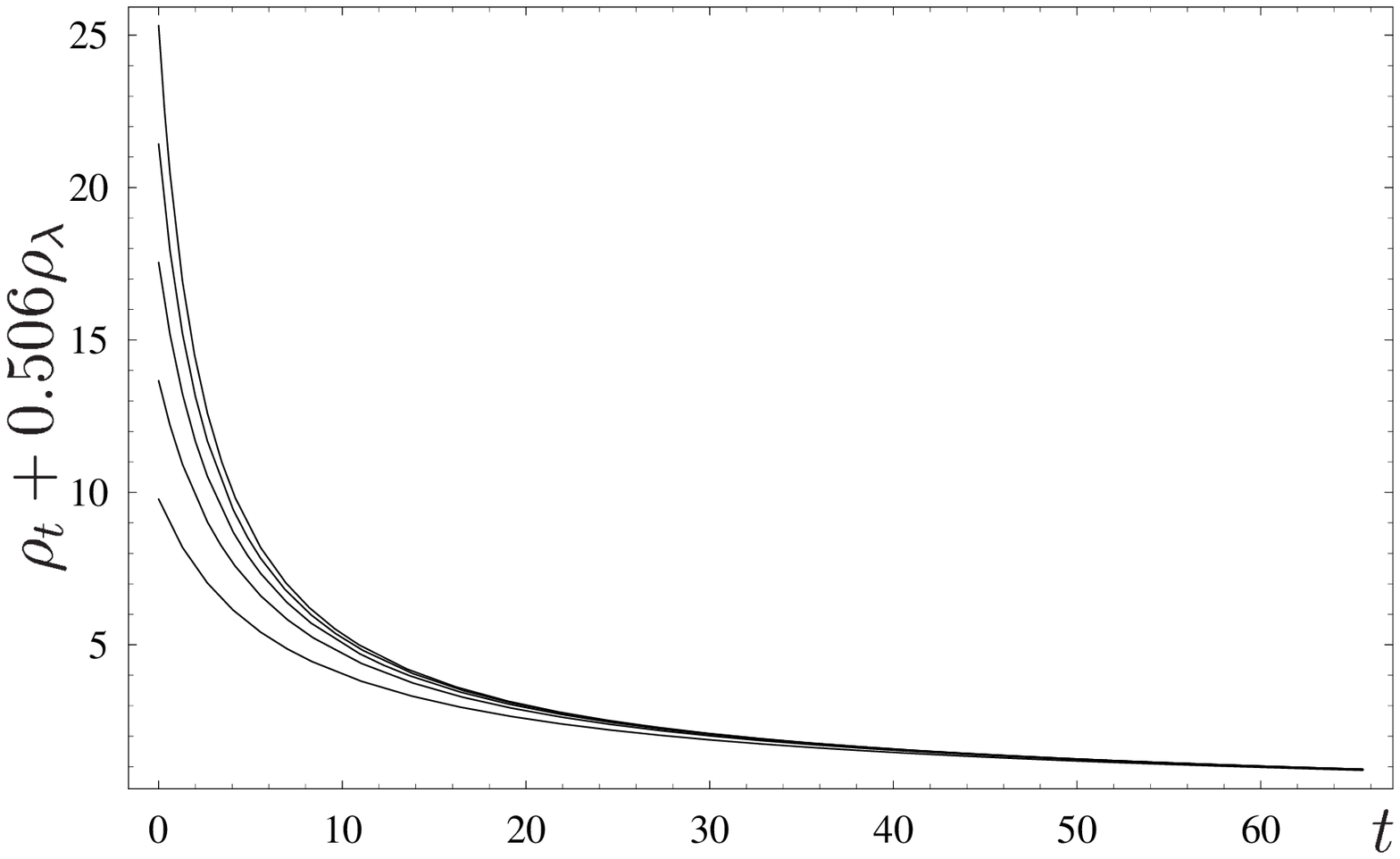}

\vspace{2mm}\hspace*{12mm}{\large\bfseries Fig.3a.}

\vspace{20mm}\includegraphics[height=96mm,keepaspectratio=true]{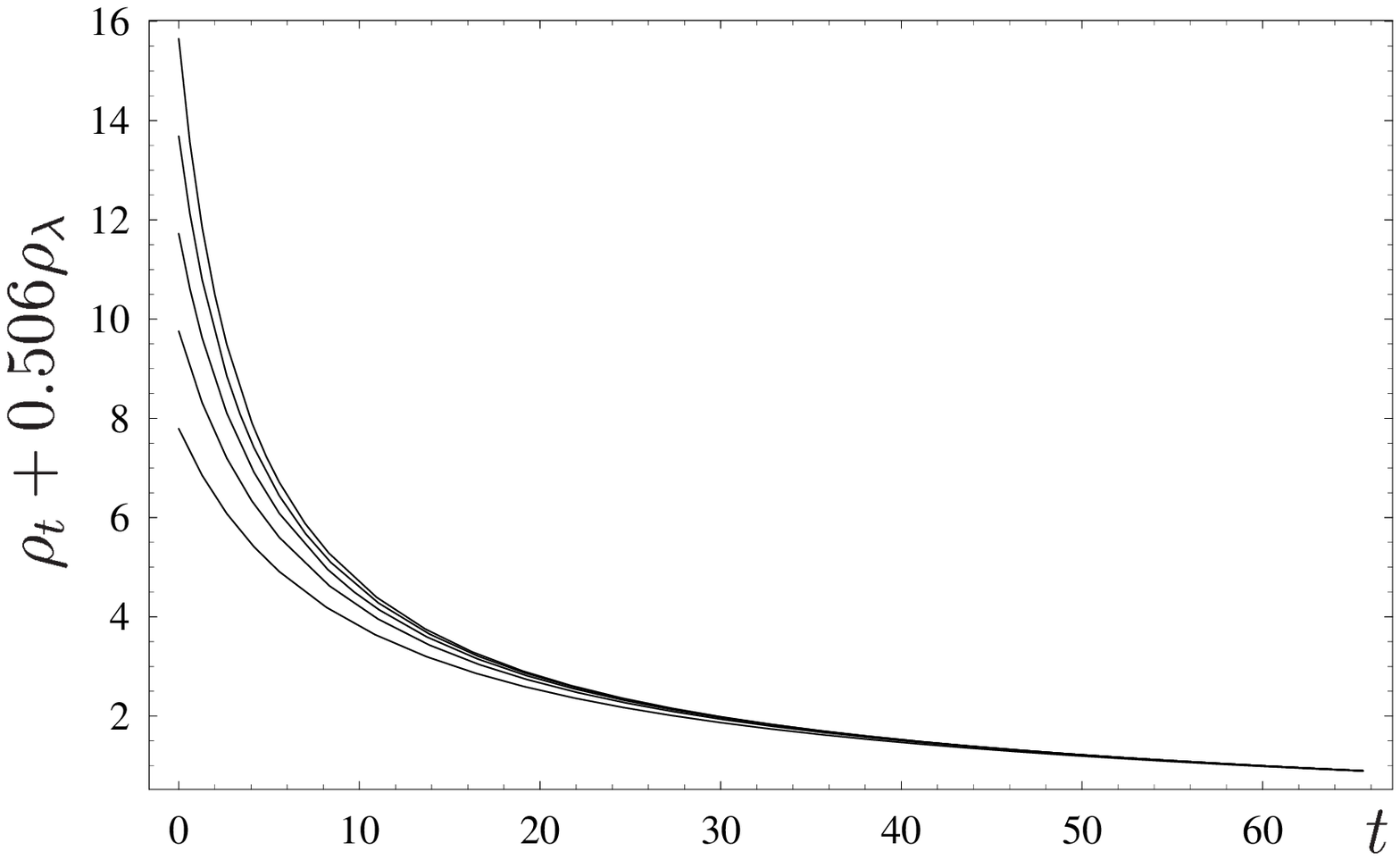}

\vspace{2mm}\hspace*{12mm}{\large\bfseries Fig.3b.}

\end{center}

\newpage

\begin{center}

\vspace*{-10mm}
\includegraphics[height=100mm,keepaspectratio=true]{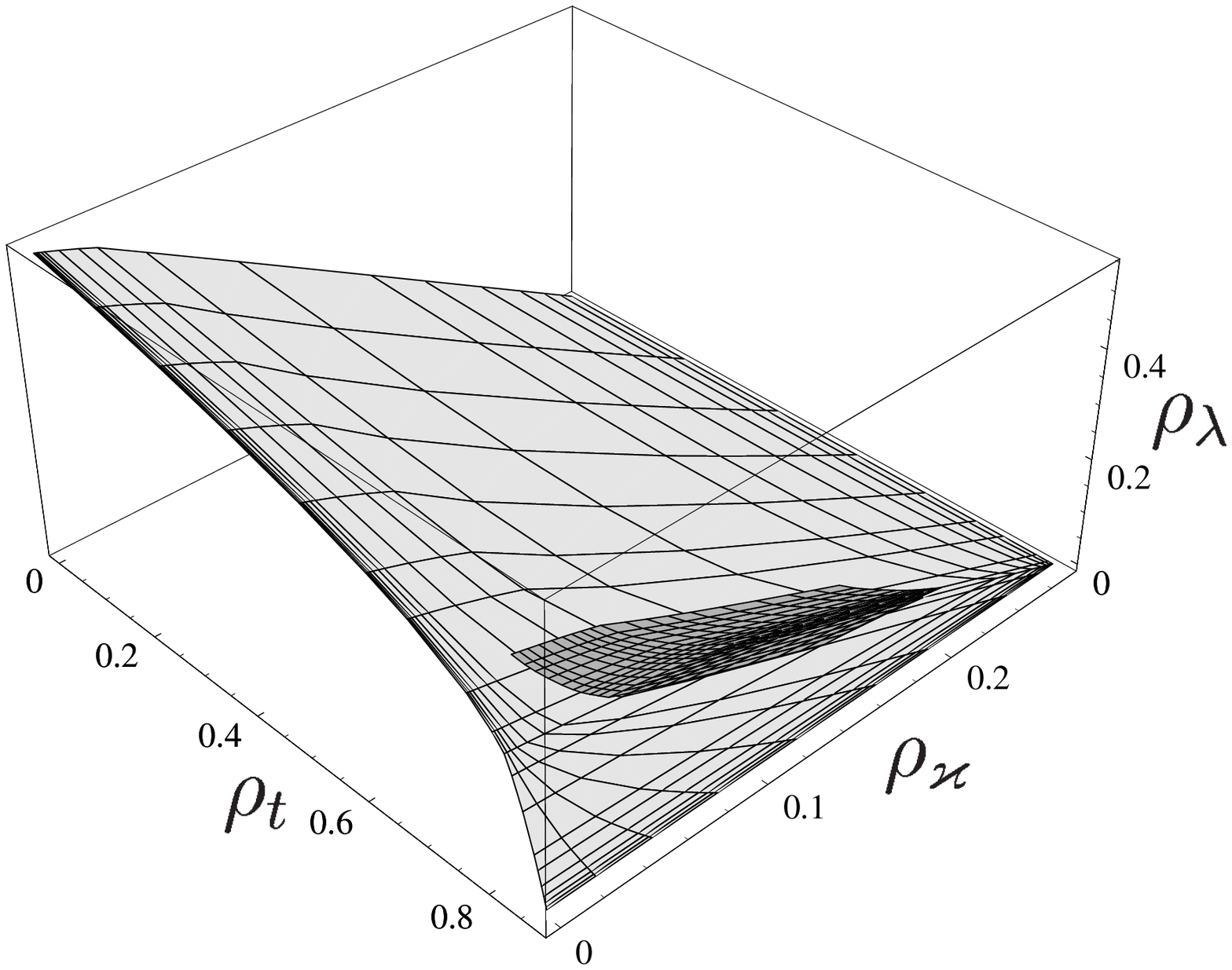}

\vspace{5mm}{\large\bfseries Fig.4a.}

\vspace{18mm}\includegraphics[height=100mm,keepaspectratio=true]{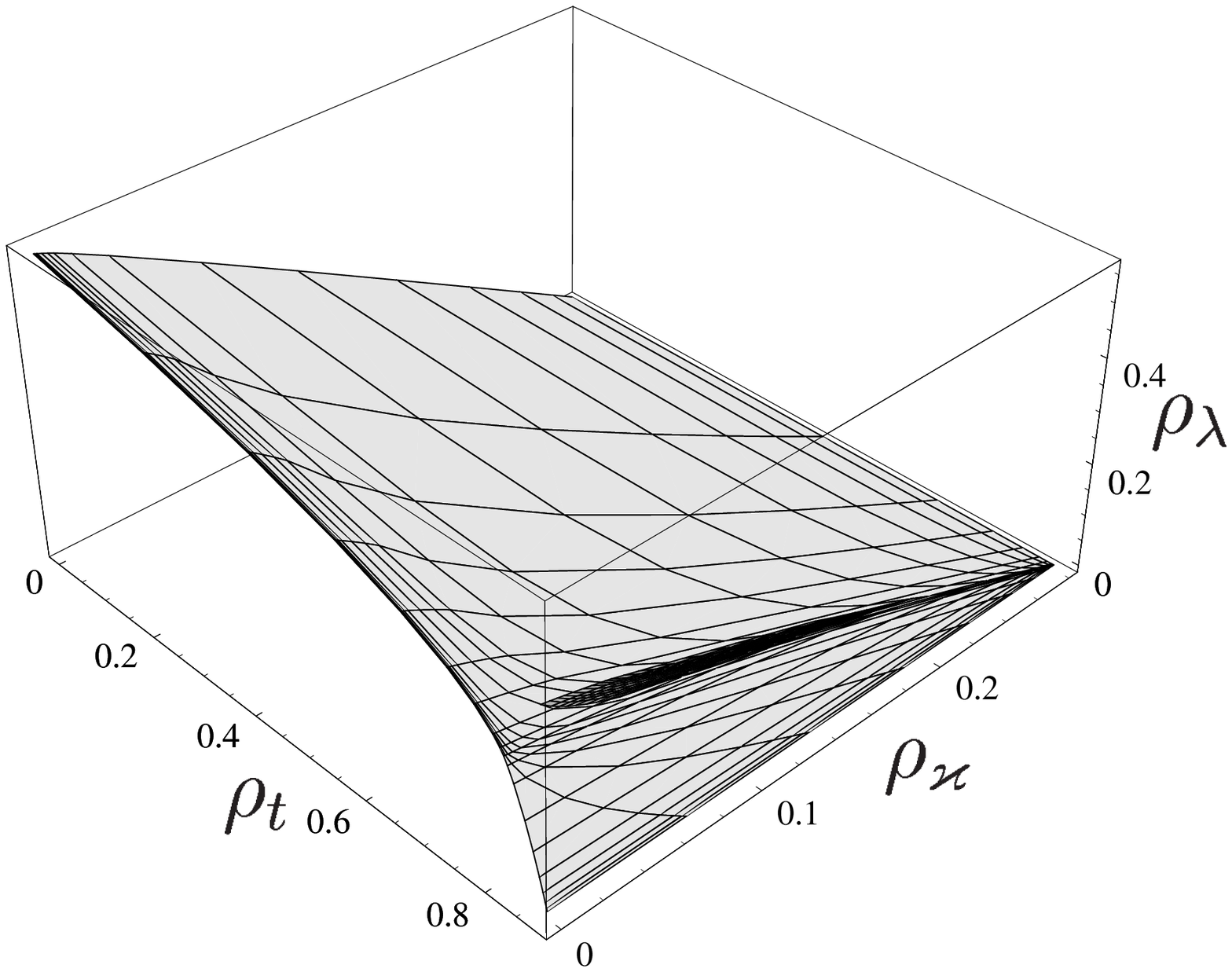}

\vspace{5mm}{\large\bfseries Fig.4b.}

\end{center}

\newpage

\begin{center}

\vspace*{-16mm}
\includegraphics[height=70mm,keepaspectratio=true]{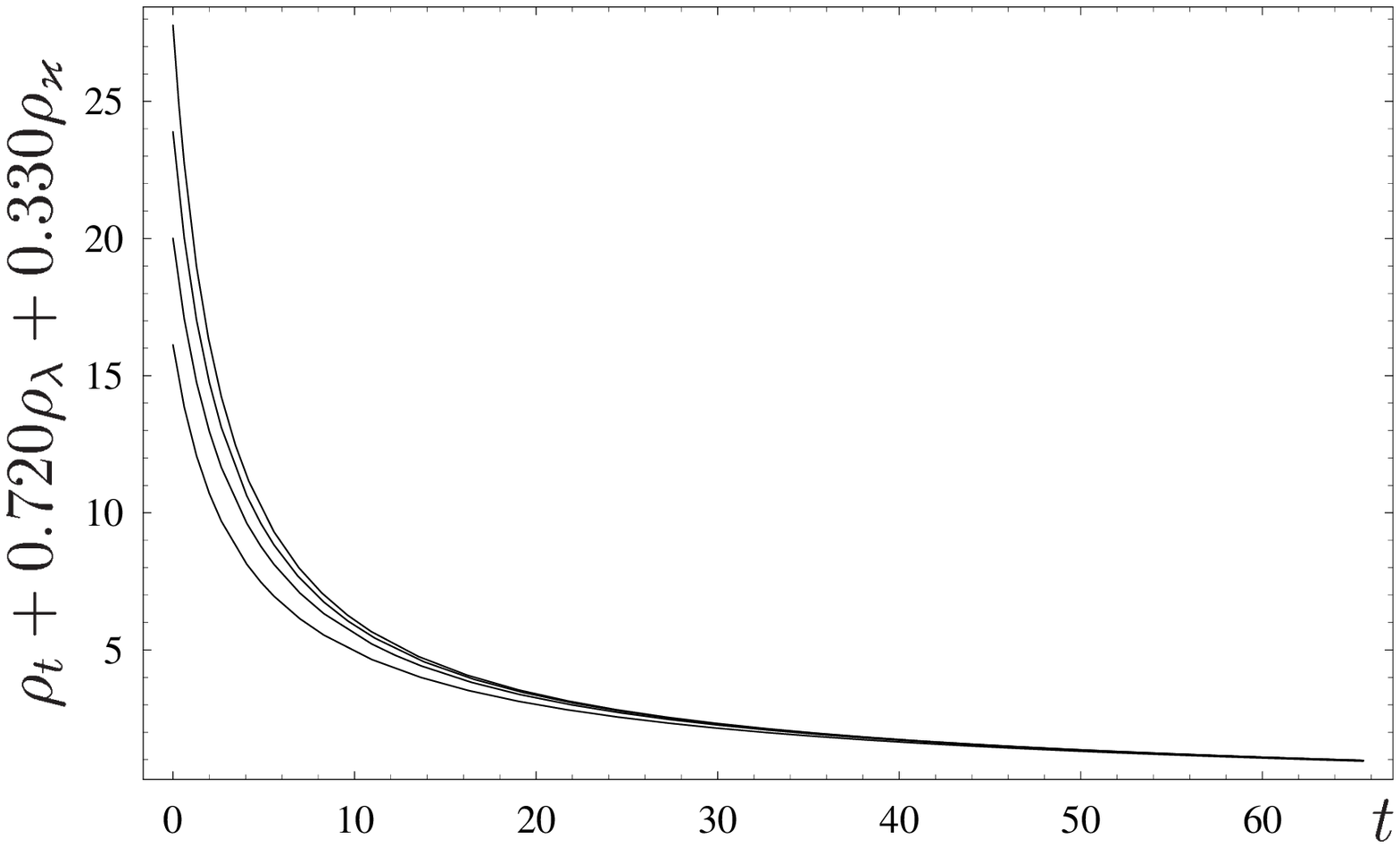}

\vspace{0mm}\hspace*{12mm}{\large\bfseries Fig.5a.}

\vspace{7mm}\includegraphics[height=70mm,keepaspectratio=true]{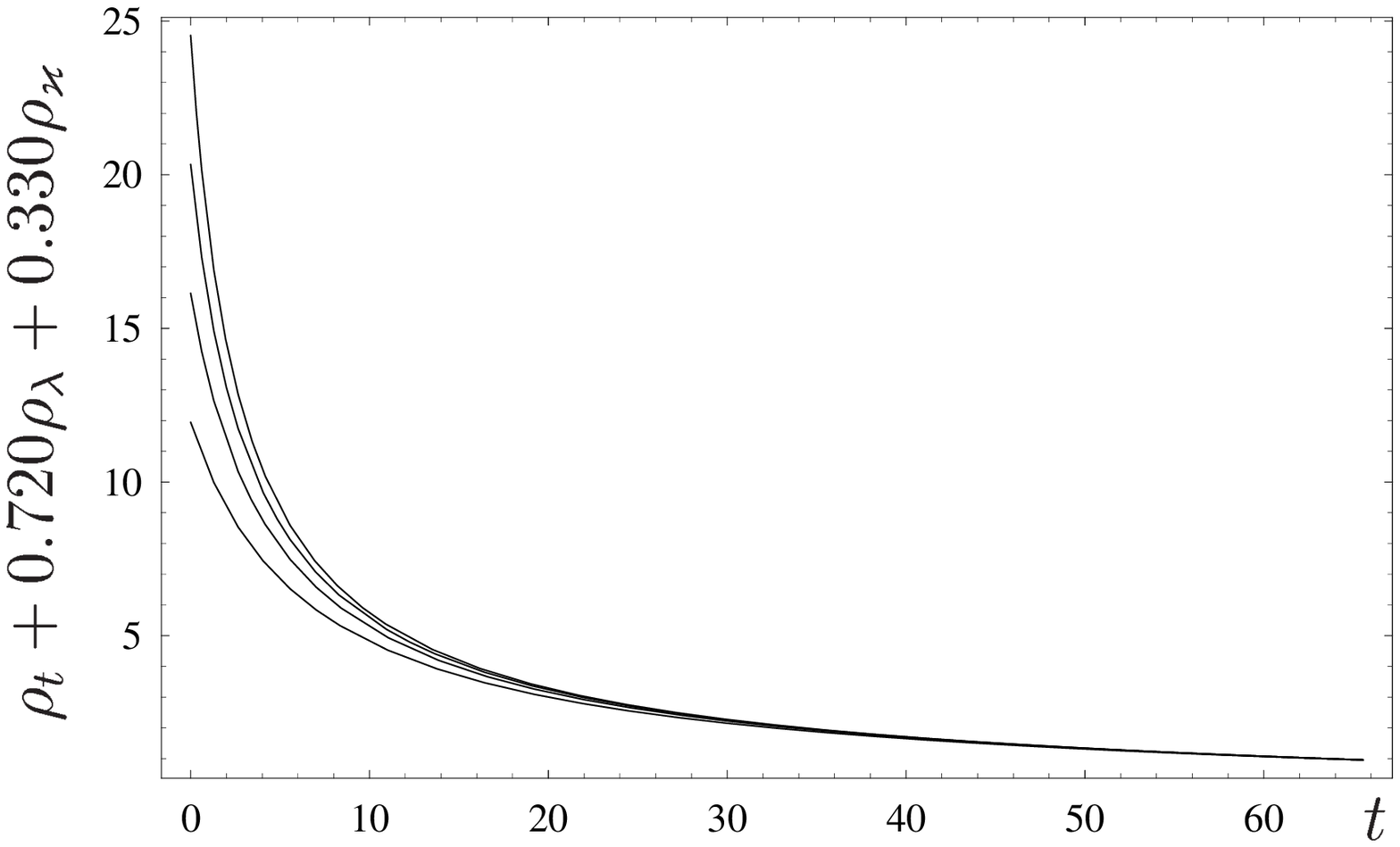}

\vspace{0mm}\hspace*{12mm}{\large\bfseries Fig.5b.}

\vspace{7mm}\includegraphics[height=70mm,keepaspectratio=true]{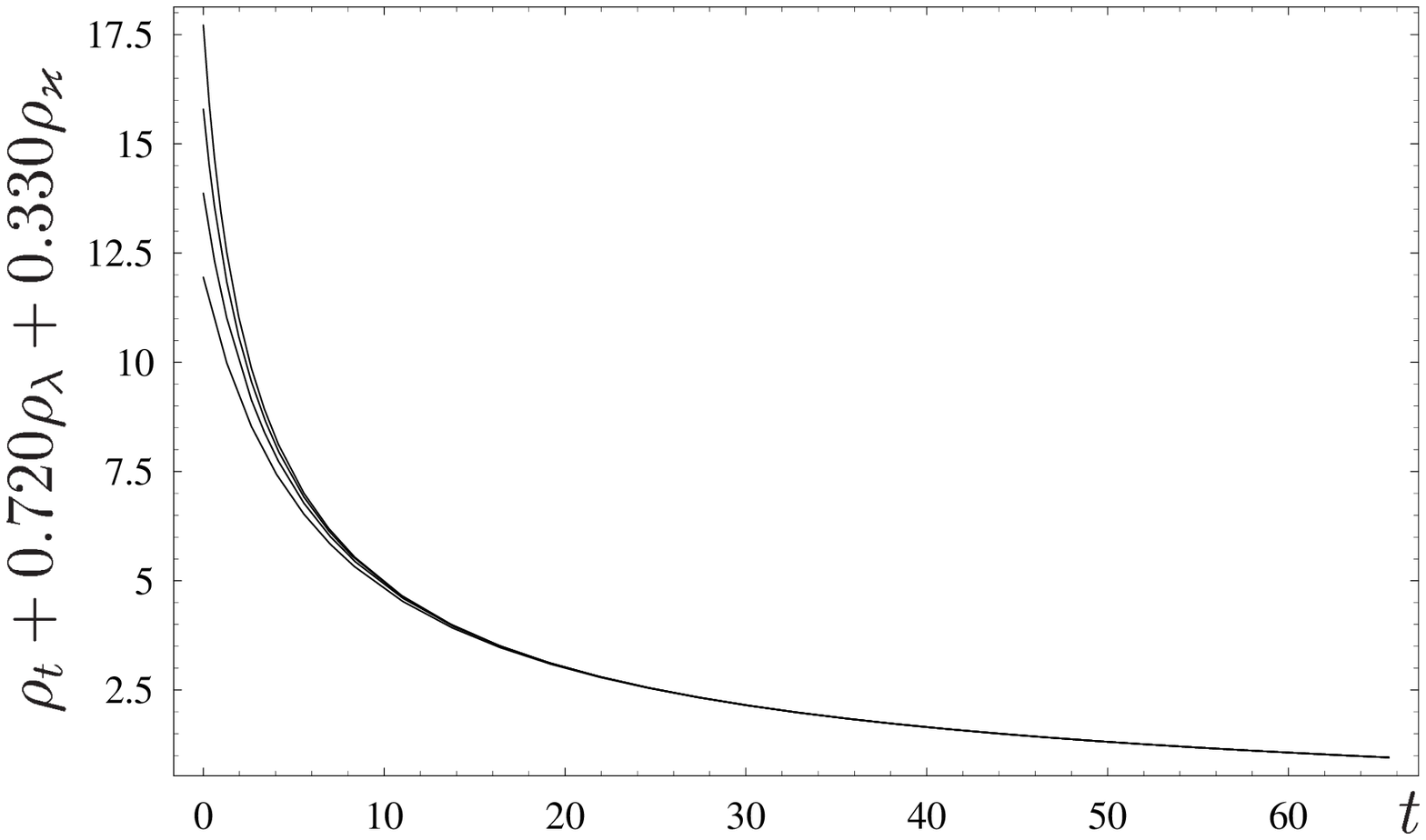}

\vspace{0mm}\hspace*{12mm}{\large\bfseries Fig.5c.}

\end{center}

\newpage

\begin{center}

\vspace*{-10mm}\includegraphics[height=100mm,keepaspectratio=true]{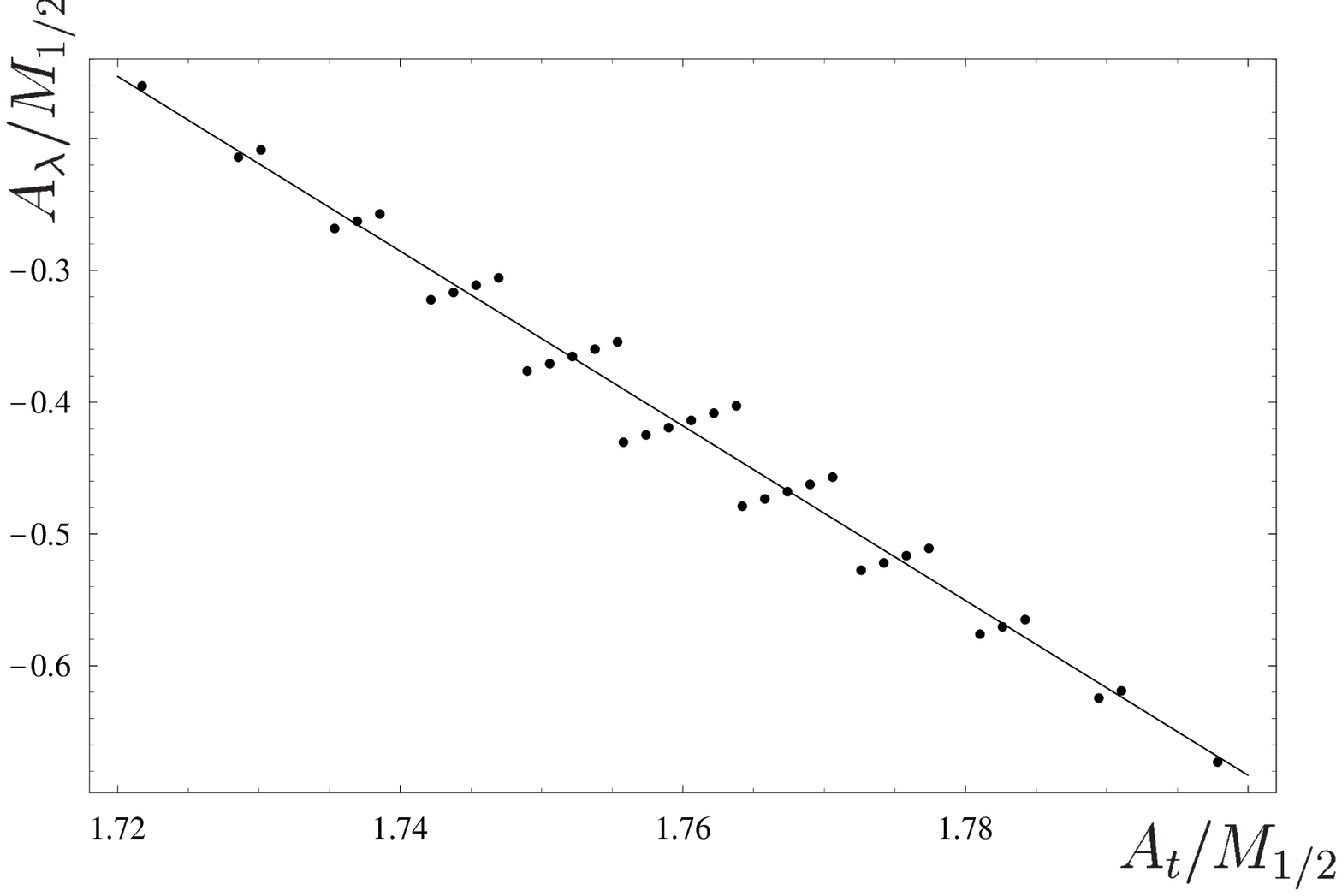}

\vspace{2mm}\hspace*{2mm}{\large\bfseries Fig.6.}

\vspace{10mm}\hspace*{4mm}\includegraphics[height=110mm,keepaspectratio=true]{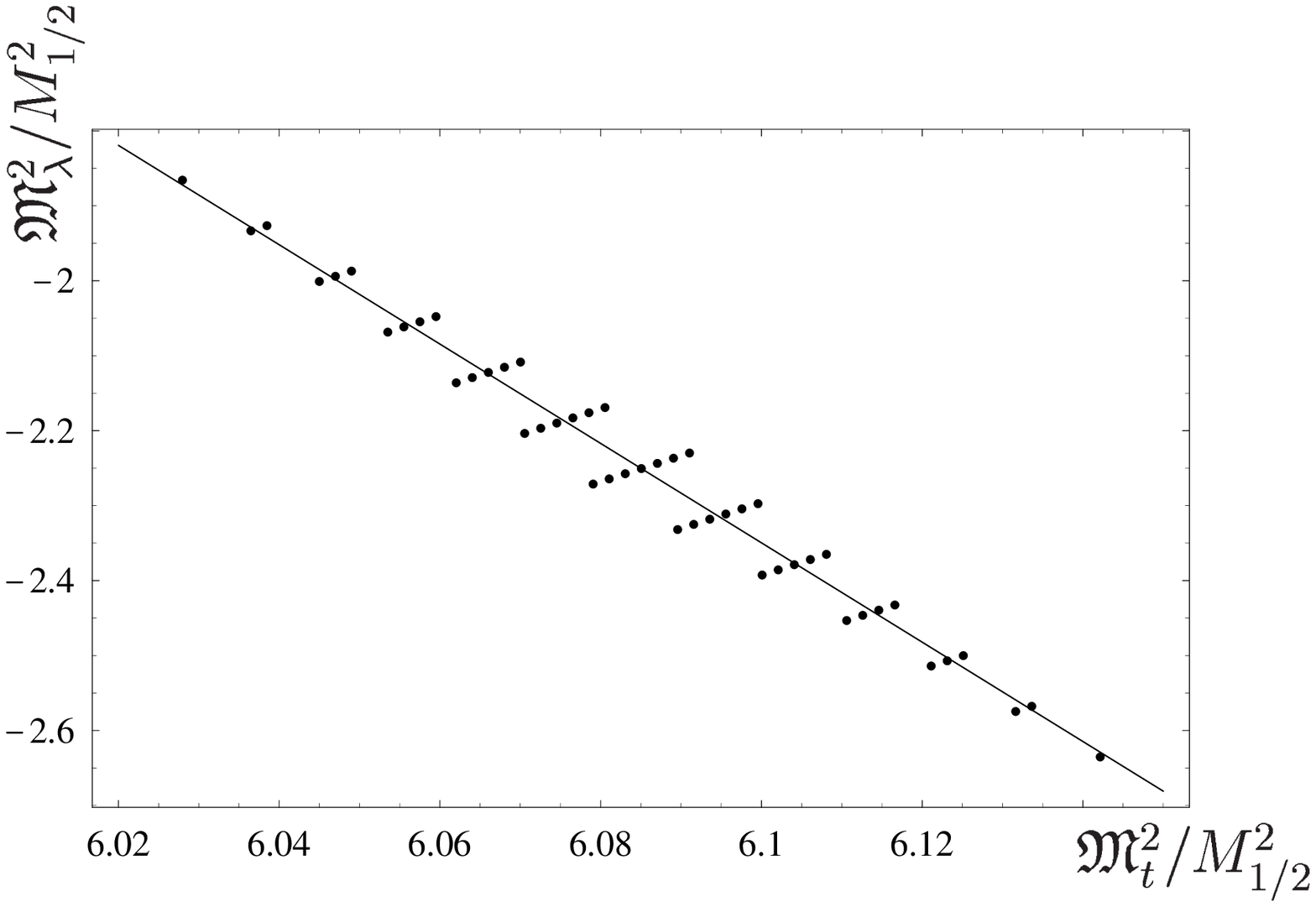}

\vspace{2mm}\hspace*{2mm}{\large\bfseries Fig.7.}

\end{center}

\newpage

\begin{center}

\vspace*{-15mm}\includegraphics[height=100mm,keepaspectratio=true]{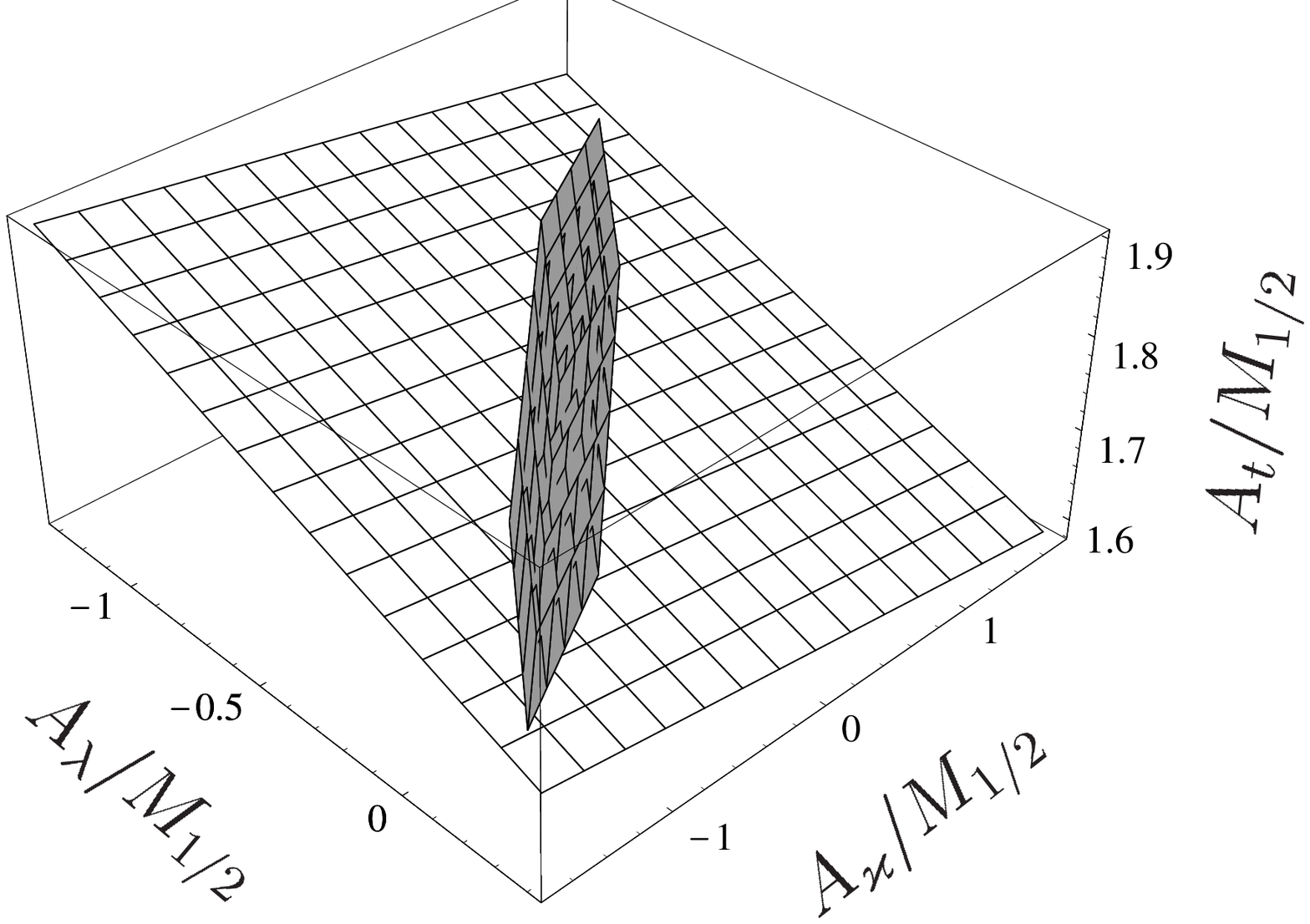}

\vspace{5mm}\hspace*{-21mm}{\large\bfseries Fig.8a.}

\vspace{15mm}\includegraphics[height=105mm,keepaspectratio=true]{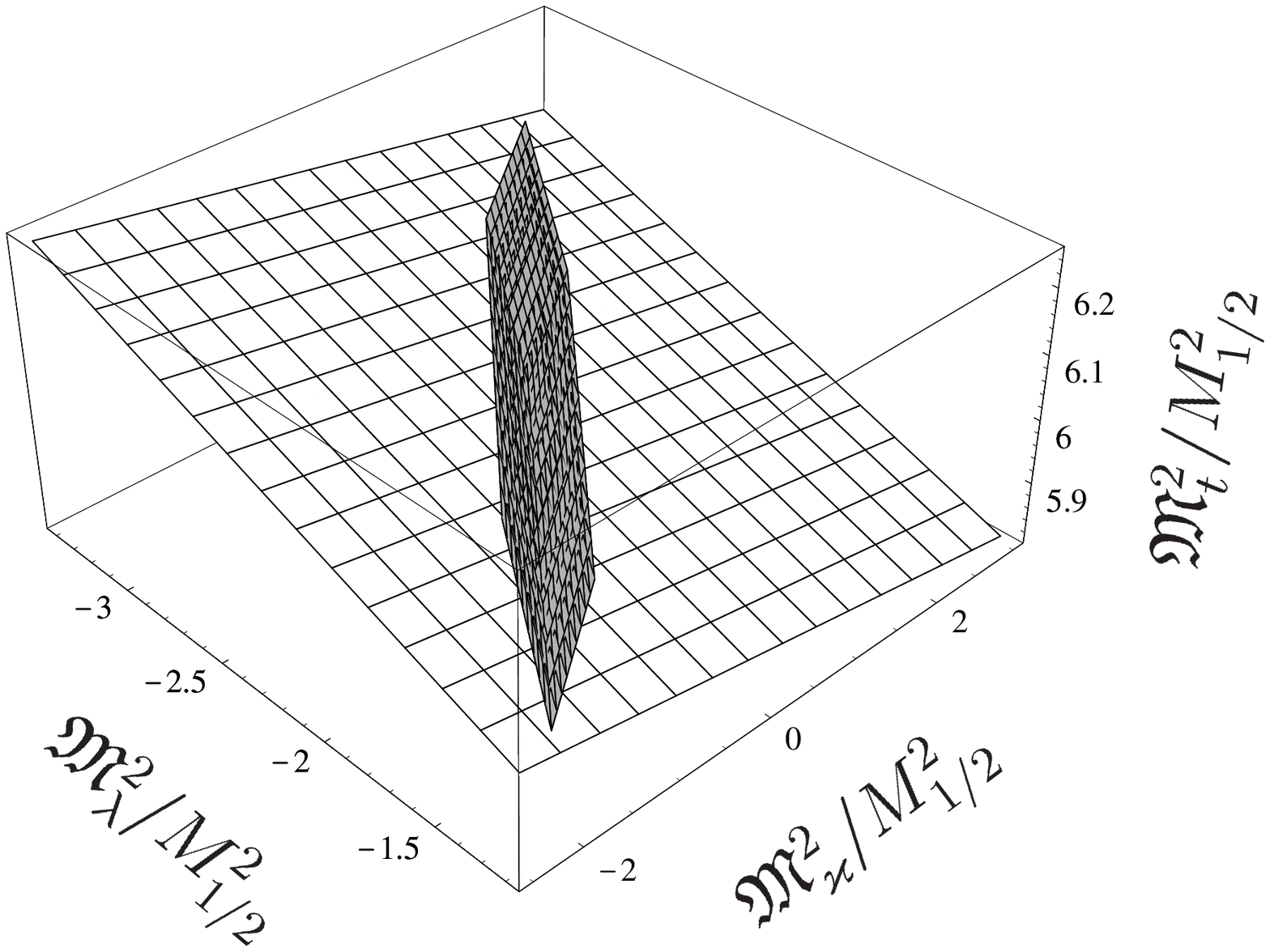}

\vspace{5mm}\hspace*{-21mm}{\large\bfseries Fig.8b.}

\end{center}

\newpage

\begin{center}

\vspace*{-15mm}\includegraphics[height=100mm,keepaspectratio=true]{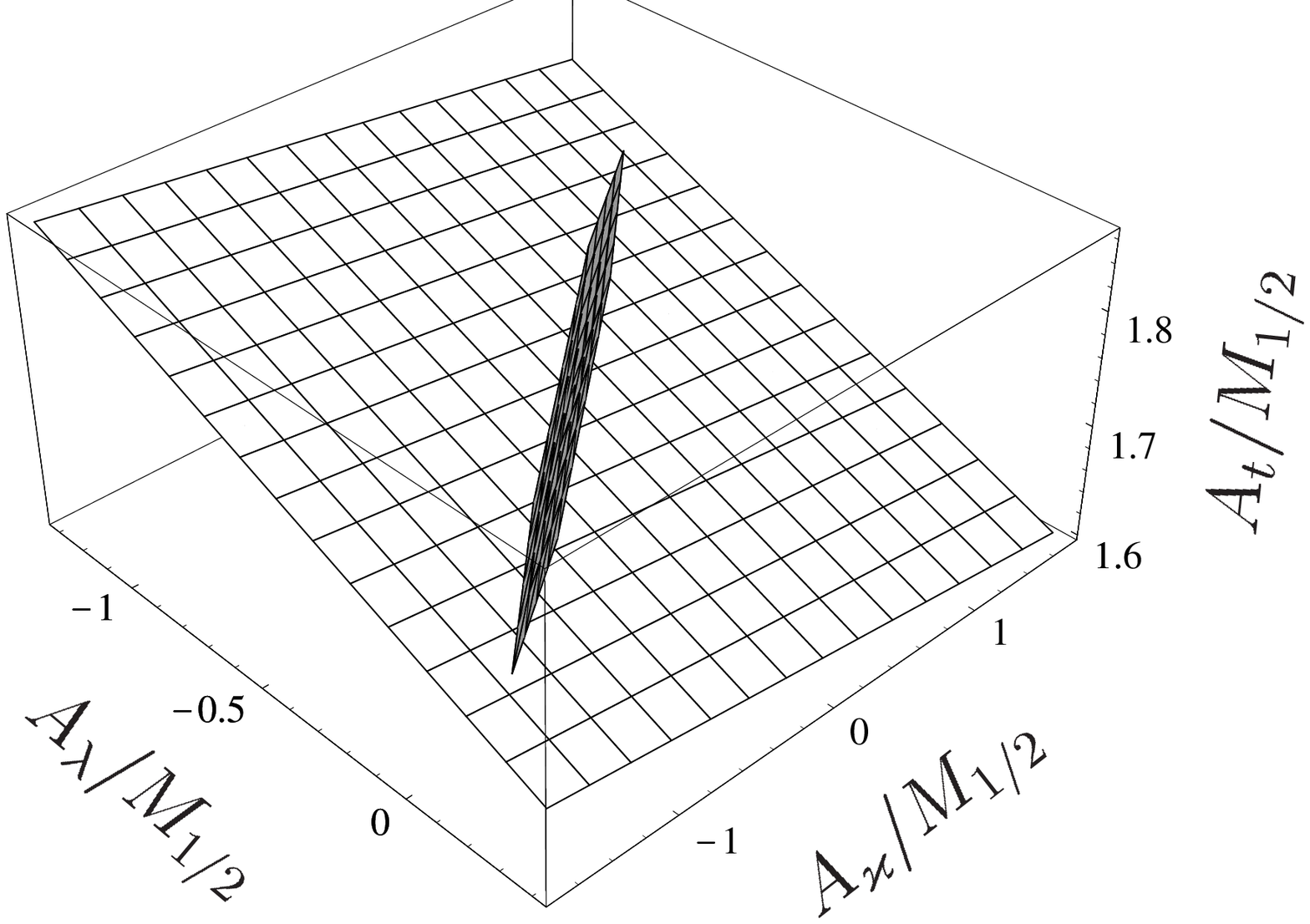}

\vspace{5mm}\hspace*{-21mm}{\large\bfseries Fig.9a.}

\vspace{15mm}\includegraphics[height=105mm,keepaspectratio=true]{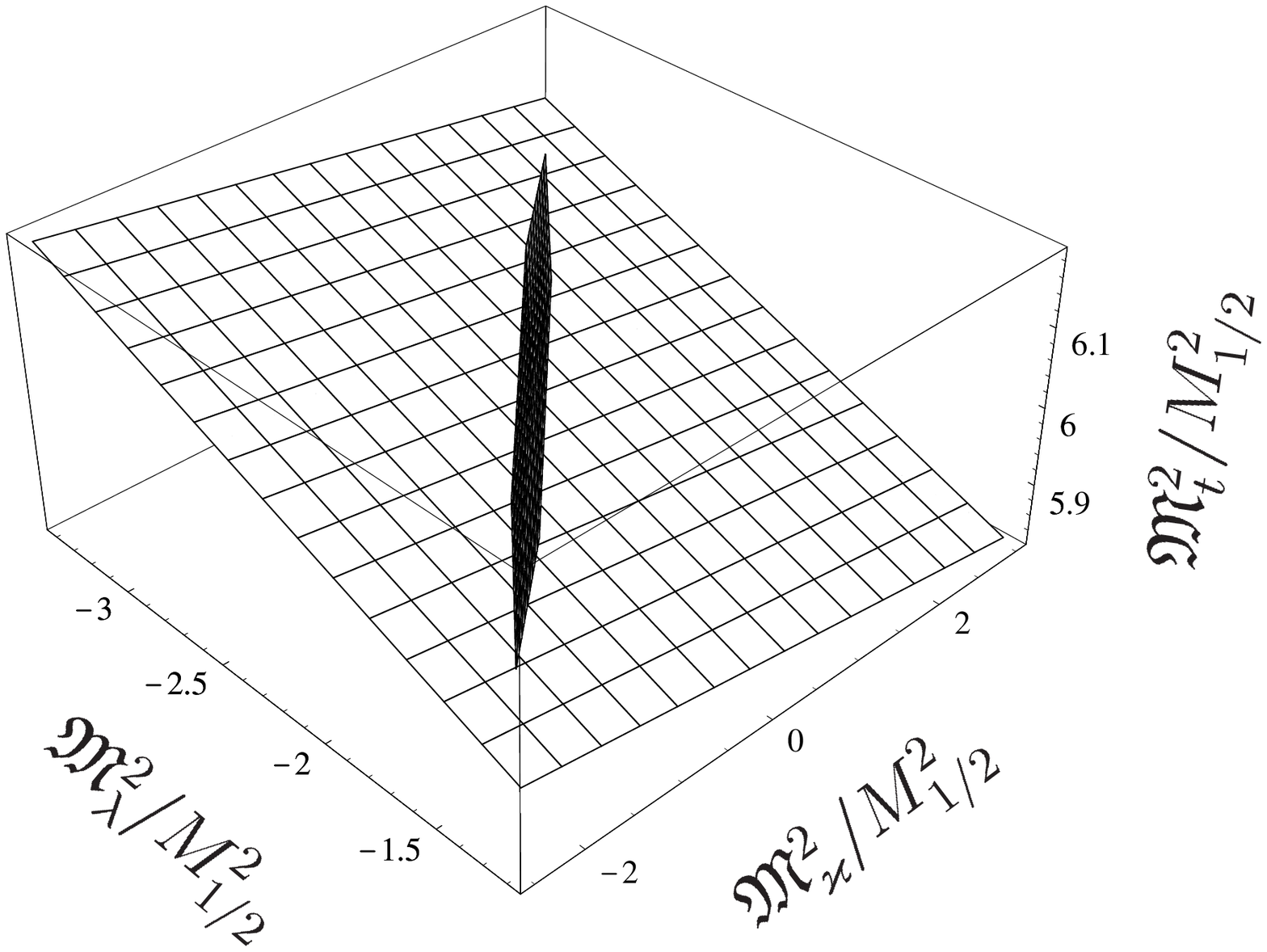}

\vspace{5mm}\hspace*{-21mm}{\large\bfseries Fig.9b.}

\end{center}

\newpage

\begin{center}

\vspace*{-15mm}\includegraphics[width=155mm,keepaspectratio=true]{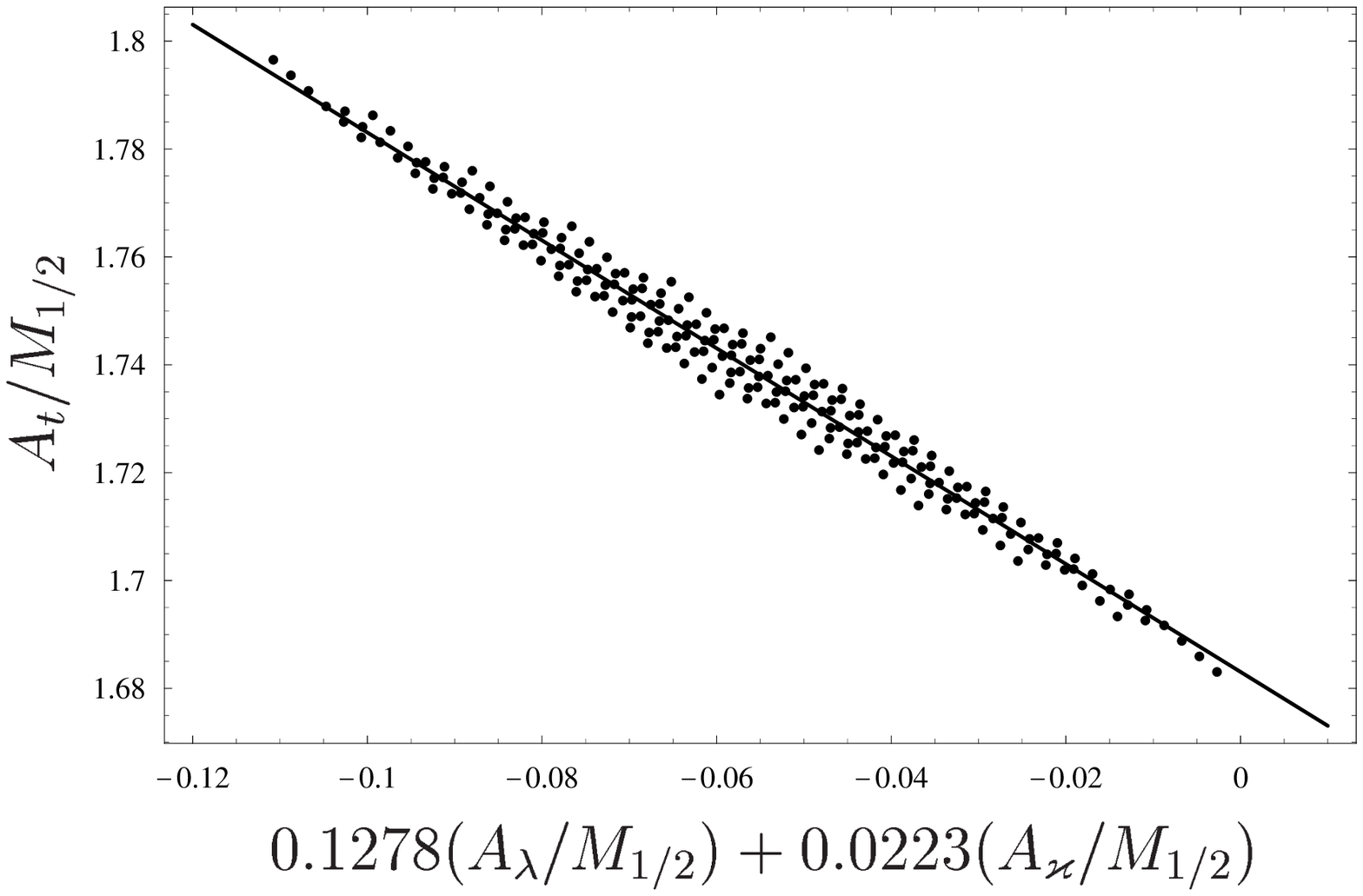}

\vspace{5mm}\hspace*{17mm}{\large\bfseries Fig.10a.}

\vspace{15mm}\includegraphics[width=155mm,keepaspectratio=true]{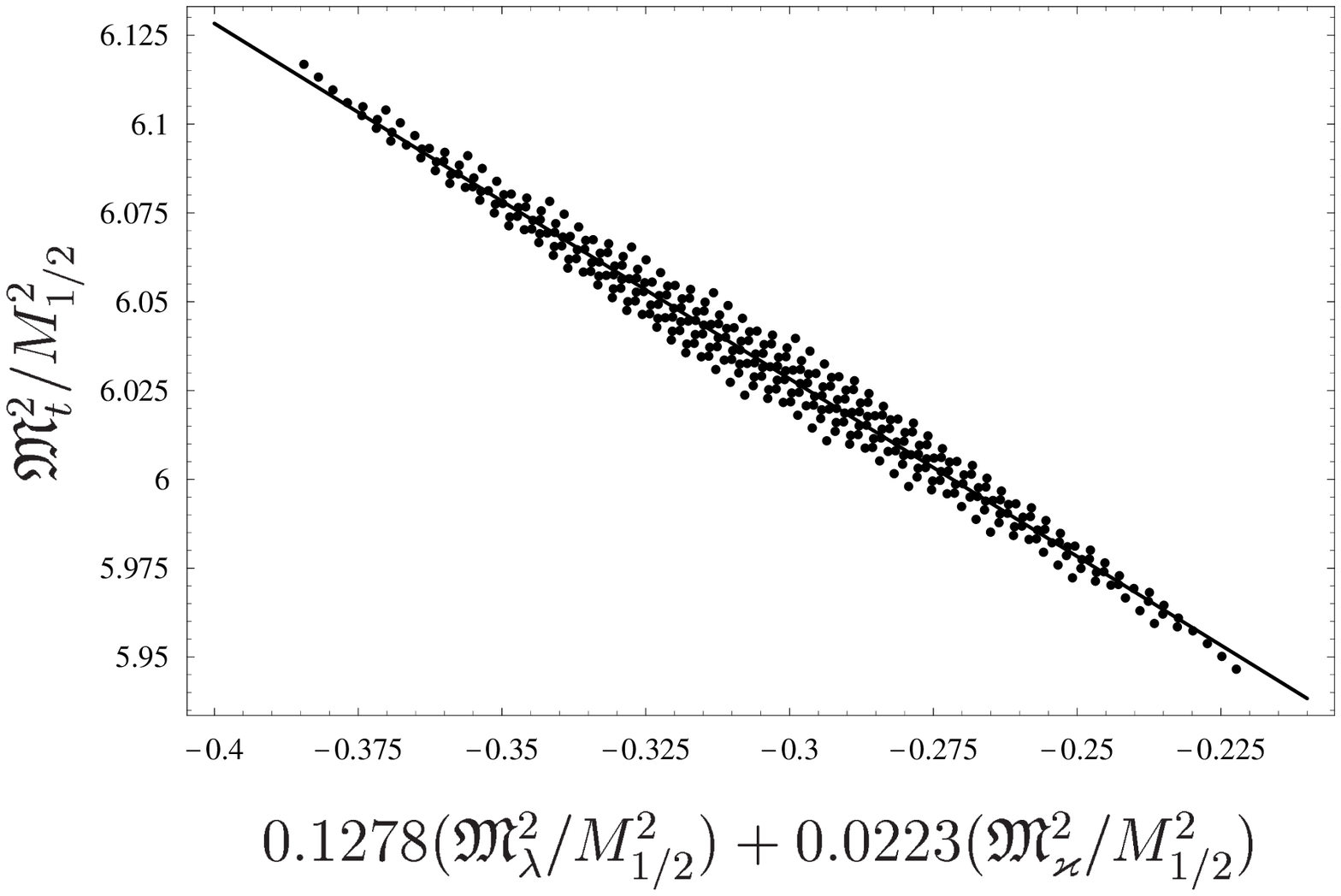}

\vspace{5mm}\hspace*{17mm}{\large\bfseries Fig.10b.}

\end{center}

\end{document}